\documentclass[oldversion]{aa}
\usepackage{txfonts}
\usepackage{natbib}
\usepackage{subfigure}
\bibpunct{(}{)}{;}{a}{}{,} 
\usepackage{graphicx}
\usepackage{amssymb}
\usepackage{threeparttable}
\usepackage{rotating}

\usepackage{epstopdf}

\begin{document}

\title{The behavior of subluminous X-ray transients near the Galactic center as observed using
the X-ray telescope aboard Swift}

\titlerunning{Swift observations of subluminous X-ray transients near the Galactic center}

\author{Nathalie Degenaar \and Rudy Wijnands}

\authorrunning{N. Degenaar \and R. Wijnands}

\institute{University of Amsterdam, Kruislaan 403, 1098 SJ, Amsterdam, the Netherlands}

\date{Received 23 July 2008 / Accepted 11 December 2008 }

\abstract {
In this paper we report on the spectral analysis of seven X-ray transients, which were found to be active during a monitoring campaign of the Galactic center carried out in 2006 and 2007 using the X-ray telescope aboard the \textit{Swift} satellite. This campaign detected new outbursts of five known X-ray transients and discovered two new systems. Their 2-10 keV peak luminosities range from $\sim 10^{34}~\mathrm{to}~6 \times 10^{36}~\mathrm{erg~s}^{-1}$, which implies that all seven X-ray transients are subluminous compared to the bright X-ray transients that have peak luminosities of $10^{37-39}$~erg~s$^{-1}$. Two of the sources discussed in this paper are confirmed neutron star systems (AX J1745.6-2901 and GRS 1741-2853), while the five others have an unknown nature. We discuss the characteristics of the observed outbursts and the duty cycles of the various systems. Several of the detected transients seem to undergo enhanced X-ray activity with levels intermediate between quiescence and full outburst. We discuss the possibility that the subluminous appearance of the eclipsing X-ray burster AX J1745.6-2901 is due to line-of-sight effects. We detected two type-I X-ray bursts with a duration of 50-60 seconds from AX J1745.6-2901, which we discuss in view of the bursting behavior of low-luminosity X-ray transients. Assuming that we are dealing with accreting neutron stars and black holes, we estimate the time-average accretion rate, $\langle \dot{M}_{\mathrm{long}} \rangle$, of the transients, which is an important input parameter for binary evolution models that attempt to explain the nature of subluminous X-ray transients. Our estimates lie in the range of $ 3 \times 10^{-13}~\mathrm{M}_{\odot}~\mathrm{yr}^{-1} \lesssim \langle \dot{M}_{\mathrm{long}} \rangle \lesssim 1 \times 10^{-10}~\mathrm{M}_{\odot}~\mathrm{yr}^{-1}$, if the systems are neutron star X-ray binaries and between $ 4 \times 10^{-14}~\mathrm{M}_{\odot}~\mathrm{yr}^{-1} \lesssim \langle \dot{M}_{\mathrm{long}} \rangle \lesssim 2 \times 10^{-11}~\mathrm{M}_{\odot}~\mathrm{yr}^{-1}$ for a scenario where the accreting object is a black hole.
Some of the systems have such low estimated mass-accretion rates that they possibly pose a challenge for binary evolution models. 
}


\maketitle

\section{Introduction}

Our Galaxy harbors many X-ray transients that spend most of their time in a dim, quiescent state, but occasionally they experience bright X-ray outbursts (typically lasting weeks to months) during which their X-ray luminosity increases by more than a factor of 100. 
Many of these transient X-ray sources can be identified with compact objects (neutron stars or black holes) accreting matter from a companion star. In such systems, the X-ray outbursts are ascribed to a sudden strong increase in the accretion rate onto the compact object.
X-ray transients can be classified based on their 2-10 keV peak luminosity\footnote{All fluxes and luminosities quoted in this paper are for the 2-10 keV energy band, unless otherwise stated.}, $L_{X}^{\mathrm{peak}}$. 
The \textit{bright} X-ray transients ($L_{X}^{\mathrm{peak}}=10^{37-39}$~erg~s$^{-1}$) have been known and extensively studied since the early days of X-ray astronomy. However, in the past decade it became clear that a group of subluminous X-ray transients ($L_{X}^{\mathrm{peak}}<10^{37}$~erg~s$^{-1}$) also exists, where the distinction is made between \textit{faint} \citep[$L_{X}^{\mathrm{peak}}=10^{36-37}$~erg~s$^{-1}$, e.g.,][]{heise99,zand01} and \textit{very-faint} \citep[$L_{X}^{\mathrm{peak}}=10^{34-36}$~erg~s$^{-1}$, e.g.,][]{sidoli99, porquet05,muno05_apj622,wijn06_monit} systems. 
Although the faint to very faint X-ray transients exhibit qualitatively different behavior than the brighter systems \citep[e.g.,][]{cornelisse02, okazaki01, king00}, this classification based on peak luminosities is not strict and hybrid systems are known to exist \citep[e.g.,][]{wijn02}. 

In particular the study of very-faint X-ray transients (VFXTs) is hampered by the sensitivity limitations of X-ray instruments, and consequently their nature is not understood well. To date, about 30 members are known, most of which are found very close to $\mathrm{Sgr~A}^{\ast}$ \citep[within $\sim$ 10 arcminutes; ][]{muno05_apj622}, but this might be a selection effect due to all the high-resolution X-ray observations in this region. Several VFXTs were found at larger distances from $\mathrm{Sgr~A}^{\ast}$ as well \citep[e.g., ][]{hands04,heinke2008}. A significant fraction ($\sim 1/3$) of the VFXTs have exhibited type-I X-ray bursts \citep[e.g., ][]{cornelisse02} and can thus be identified with neutron stars accreting matter from, most likely, a low-mass (i.e., $M \lesssim 1 \mathrm{M_{\odot}}$) companion. The low outburst luminosities characteristic of VFXTs combined with what is known about their duty cycles, imply that these low-mass X-ray binaries (LMXBs) have very low time-averaged mass accretion rates, which could challenge our understanding of their evolution \citep{king_wijn06}. 

There might also be other types of sources that can produce subluminous X-ray outbursts. It is conceivable that some systems are compact objects that are transiently accreting at a very low level from the strong stellar wind of a high-mass star or the circumstellar matter around a Be star \citep[e.g.,][]{okazaki01}. In addition, some strongly magnetized neutron stars ($B \sim10^{14}-10^{15}$~G, magnetars) are observed to experience occasional X-ray outbursts with peak luminosities of $\sim10^{35}$~erg~s$^{-1}$ \citep{ibrahim04,muno07_magnetar} and can thus be classified as VFXTs. The cause of their outbursts is unknown, but is likely related to magnetic field decay \citep[e.g.,][]{ibrahim04}. Furthermore, \citet{mukai08} recently pointed out that classical novae can be visible as 2-10 keV X-ray sources with luminosities in the range of a few times $10^{34-35}~\mathrm{erg~s}^{-1}$ for weeks to months \citep[see Fig.~1 of][]{mukai08}. The X-ray emission is thought to emerge from shocks within the matter that is ejected during the nova. 

Here we present the analysis of seven X-ray transients, that were found active during a monitoring campaign of the Galactic center (GC) by the X-ray telescope (XRT) aboard the \textit{Swift} satellite \citep{kennea_monit}, carried out in 2006 and 2007.

\section{Observations and data analysis}\label{obs_ana}

The GC was monitored almost daily with the XRT aboard \textit{Swift}, from February 24, 2006, until November 2, 2007\footnote{The campaign continues in 2008, but a detailed  discussion of the 2008 data is beyond the scope of this paper.}, with exclusion of the epochs from November 3, 2006, till March 6, 2007 (due to Solar constraints) and August 11 till September 26, 2007 \citep[due to a safe-hold event;][]{swift_offline07}.  Each \textit{Swift}/XRT pointing typically lasted $\sim 1$ ksec, although occasionally longer exposures (up to $\sim 13$ ksec) were carried out. Most of the data was collected in Photon Counting (PC) mode, albeit sometimes an unusual high count rate (due to the occurrence of a type-I X-ray burst) induced an automated switch to the Windowed Timing (WT) mode. 
We obtained all observations of 2006-2007 GC monitoring campaign from the \textit{Swift} data archive. 

The XRT data were processed with the task \texttt{xrtpipeline} using standard quality cuts and event grades 0-12 in PC mode (0-2 in WT mode)\footnote{See http://heasarc.gsfc.nasa.gov/docs/swift/analysis for standard \textit{Swift} analysis threads}. We searched the data for transient X-ray sources by comparing small segments of \textit{Swift} data, spanning $\sim 5$ ksec, with one another. We found a total of seven different X-ray transients with peak luminosities $ \gtrsim 10^{34}~\mathrm{erg~s}^{-1}$. The source coordinates and associated uncertainties of the detected transients were found by running the XRT software task \texttt{xrtcentroid} on the data. The results are listed in Table~\ref{tab:vfxts}.
A source was considered in quiescence when it was not detected within a databin of approximately 5 ksec by visual inspection. The unabsorbed 2-10 keV flux corresponding to this threshold depends on the assumed spectral model, but is roughly $2 \times 10^{-13}~\mathrm{erg~cm}^{-2}~\mathrm{s}^{-1}$. This translates into a luminosity of $\sim 1.5 \times 10^{33}~\mathrm{erg~s}^{-1}$ for a distance of 8 kpc.

\begin{figure*}[tb]
 \begin{center}
 \subfigure[]{\label{fig:ds9-a}\includegraphics[width=9.0cm]{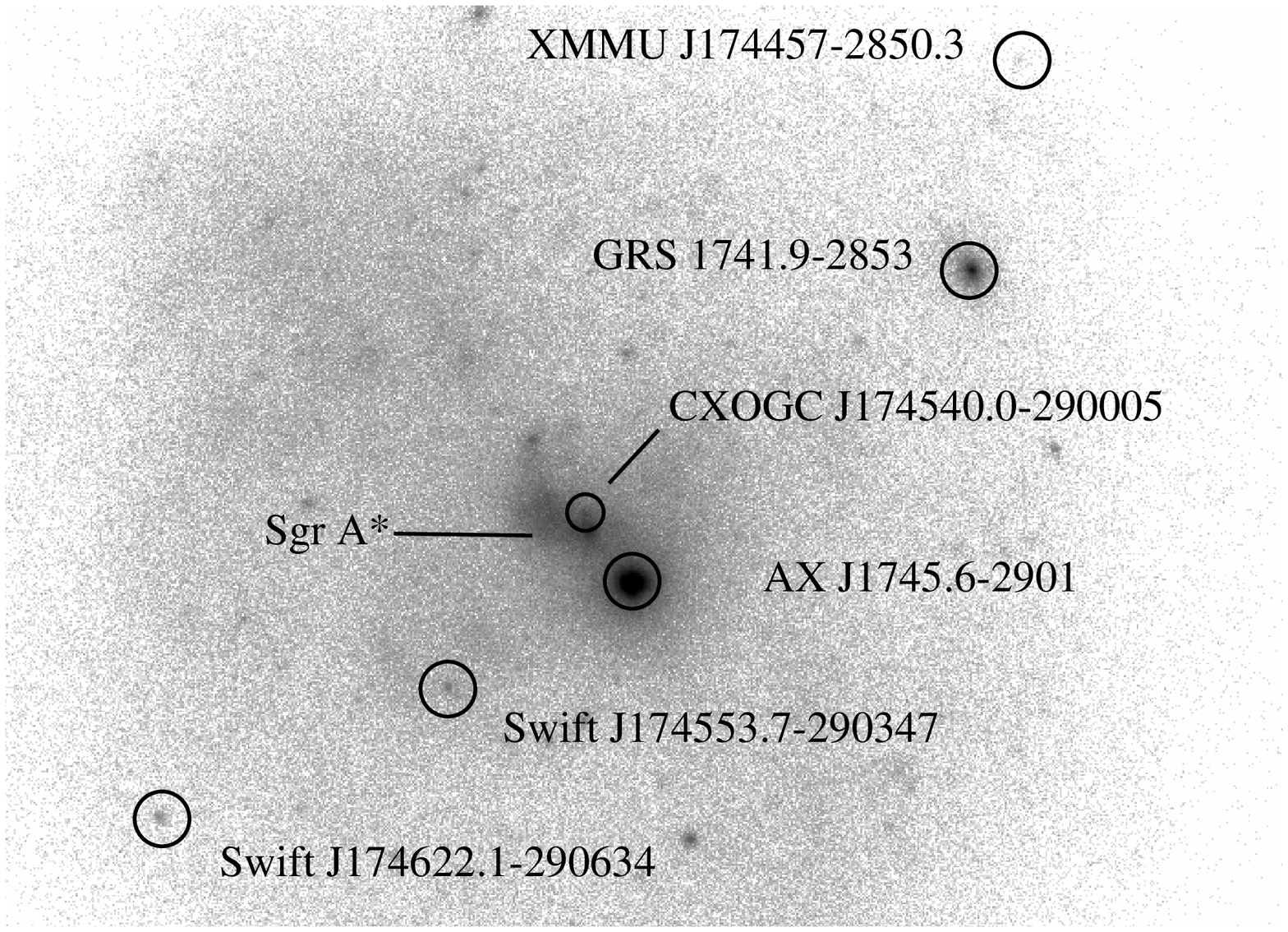}}
 \subfigure[]{\label{fig:ds9-b}\includegraphics[width=9.0cm]{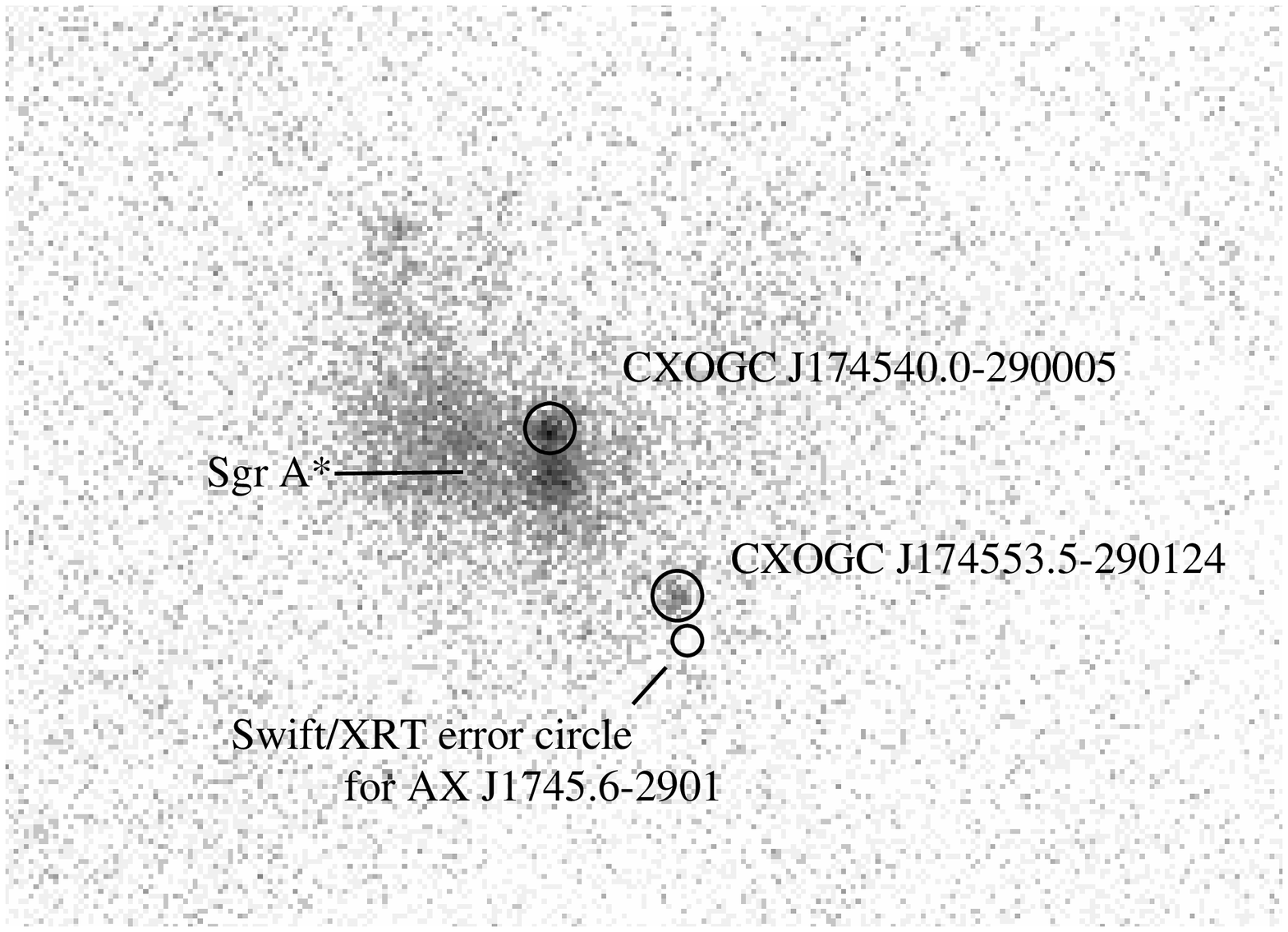}}
    \end{center}
\caption[]{X-ray images (0.3-10 keV) of the GC obtained with \textit{Swift}/XRT (North is up and East is to the right). a) Merged image of all PC mode observations carried out in 2006 and 2007. The known X-ray transients AX J1745.6-2901, CXOGC J174540.0-290005, GRS 1741.9-2853 and XMM J174457-2850.3, as well as the newly discovered subluminous X-ray transients Swift J174553.7-290347 and Swift J174622.1-290634 can be seen in this image. b) Zoomed image of the inner region around  $\mathrm{Sgr~A}^{\ast}$ of the epoch June-November 2006, during which CXOGC J174535.5-290124 and CXOGC J174540.0-290005 were both detected in an active state. The \textit{Swift}/XRT position for AX J1745.6-2901 is also plotted, to show that the active object does not coincide with the coordinates of AX J1745.6-2901 and is in fact a distinct source.}
  \end{figure*}

Figure~1 shows two 0.3-10 keV images of the \textit{Swift}/XRT campaign, which covered a total field of $\sim 26'~\mathrm{x} ~26'$ of sky around $\mathrm{Sgr~A}^{\ast}$ (note that individual pointings have a smaller field of view, FOV). Figure~\ref{fig:ds9-a} displays a merged image of all PC mode observations carried out in 2006 and 2007. Apart from many persistent X-ray sources and strong diffuse emission around $\mathrm{Sgr~A}^{\ast}$, Fig.~\ref{fig:ds9-a} shows six different X-ray transients with peak luminosities $ \gtrsim 10^{34}~\mathrm{erg~s}^{-1}$ (listed in Table~\ref{tab:vfxts}). Figure~\ref{fig:ds9-b} is a zoomed image of the inner region around $\mathrm{Sgr~A}^{\ast}$, taken from the epoch June 30 till November 2, 2006. This was the only episode during the entire 2006-2007 \textit{Swift} monitoring campaign in which AX J1745.6-2901 was not active, and a seventh active transient, CXOGC J174535.5-290124, could be detected. CXOGC J174535.5-290124 and AX J1745.6-2901 are so close together, that \textit{Swift} cannot spatially resolve both sources when the latter, which is the brighter of the two, is active. Apart from CXOGC J174535.5-290124, Fig.~\ref{fig:ds9-b} also shows CXOGC J174540.0-290005, which lies North of $\mathrm{Sgr~A}^{\ast}$.

We extracted source lightcurves and spectra (using \textit{XSelect} version 2.3) from the event lists using a circular region with a radius of 10 or 15 pixels (the largest regions were used for the brightest sources). Spectra were extracted only from the data in which a source was active, whereas lightcurves were constructed from all data where a source was in FOV. Corresponding background lightcurves and spectra were averaged over a set of three nearby source-free regions, each of which had the same shape and size as the source region. For none of the transients it was possible to use an annulus for the background subtraction, either because the objects were too close to the edge of the CCD or using an annular background region would encompass too much contamination from nearby X-ray sources or diffuse emission around $\mathrm{Sgr~A}^{\ast}$. The spectra were grouped using the FTOOL \texttt{grppha}, to contain bins with a minimum number of 20 photons.

Following event selection, exposure maps were generated with \texttt{xrtexpomap} to correct the spectra for fractional exposure loss due to bad columns on the CCD \citep{abbey06}\footnote{See also http://www.swift.ac.uk/XRT.shtml.\label{foot:expo}}. The generated exposure maps were used as input to create ancillary response files (ARF) with \texttt{xrtmkarf}. We used the latest versions of the response matrix files (v10; RMF) from the CALDB database. 
For the brightest of the seven transients, AX J1745.6-2901 and GRS 1741.9-2853, the 2007 PC mode data was affected by pile-up. We attempted to correct for the consequent effect on spectral shape and loss in source flux by the same methods as described by \citet{vaughan06}\footnote{See also http://www.swift.ac.uk/pileup.shtml.\label{foot:pileup}}. 

Using \textit{XSPEC} \citep[version 11.1;][]{xspec}, we fitted all grouped spectra with a powerlaw continuum model modified by absorption. From these fits we deduce the 2-10 keV mean unabsorbed outburst flux for each source and combined this with the average 2-10 keV \textit{Swift}/XRT count rate of the outburst to infer a flux-to-count rate conversion factor. This factor was then used to determine the 2-10 keV unabsorbed peak flux for each source from the maximum count rate observed.

\begin{table*}
\begin{threeparttable}[tb]
\begin{center}
\caption{\textit{Swift}/XRT positions (J2000) and errors of the transients found active during the GC monitoring campaign of 2006-2007.}
\begin{tabular}{l l l l l l}
\hline \hline
\ & \multicolumn{3}{c}{Coordinates\tnote{a}}  \\
\cline{2-4}
Name & R.A. & Decl. & Err. & Comments/Association & References\tnote{b} \\
\ & (h m s) & $(^{\circ}~'~'')$ & ($''$) & \ & \  \\
\hline
AX J1745.6-2901 & 17:45:35.44 & $-$29:01:33.6 & 3.5 & Swift J174535.5-290135/CXOGC J174535.6-290133 & 1,2,3,4\\
CXOGC J174535.5-290124 & 17:45:35.80 & $-$29:01:21.0 & 3.5 & New outburst from known X-ray transient & 4,5 \\
CXOGC J174540.0-290005 & 17:45:40.29 & $-$29:00:05.4 & 3.5 & Swift J174540.2-290005 & 6,7,8,9\\
Swift J174553.7-290347 & 17:45:53.79 & $-$29:03:47.8 & 3.5 & New X-ray transient, CXOGC J174553.8-290346? & This work, 4\\
Swift J174622.1-290634 & 17:46:22.14 & $-$29:06:34.7  & 3.6 & New X-ray transient & This work  \\
GRS 1741.9-2853 &  17:45:02.43 & $-$28:54:50.0 & 3.5 &New outburst from known X-ray transient & 2,10,11,12, 14\\
XMM J174457-2850.3 &  17:44:57.30 & $-$28:50:20.8 & 4.0 & New outburst from known X-ray transient & 12, 13, 14\\
\hline
\end{tabular}
\label{tab:vfxts}
\begin{tablenotes}
\item[a] The quoted coordinate errors refer to $90\%$ confidence level and were calculated using the software tool \texttt{xrtcentroid}.
\item[b]References: 1=\citet{kennea06_atel753}, 2=\citet{porquet07}, 3=\citet{maeda1996}, 4=\citet{muno04_apj613}, 5=\citet{wijn05_atel638}, 6=\citet{kennea06_atel920}, 7=\citet{kennea06_atel921}, 8=\citet{wang06_atel935}, 9=\citet{muno05_apj622}, 10=\citet{muno03_grs}, 11=\citet{wijnands07_atel1006}, 12=\citet{wijn06_monit}, 13=\citet{sakano05}, 14=\citet{muno07_atel1013}.
\end{tablenotes}
\end{center}
\end{threeparttable}
\end{table*}

\subsection{Chandra data}
To obtain more accurate position information for the X-ray transients, we searched for \textit{Chandra} archival data of the time the transients were in outburst. We found several \textit{Chandra} observations at times when our seven \textit{Swift} transients were active (see Table~\ref{tab:chandra}). We analyzed these \textit{Chandra} data using the CIAO tools (version 4.0) and the standard \textit{Chandra} analysis threads\footnote{Listed at http://asc.harvard.edu.}. The \textit{Chandra} source positions and associated errors were determined using the tool \texttt{wavdetect} and are also listed in Table~\ref{tab:chandra}.

\subsection{Time-averaged accretion rates}\label{subsec:accrates}
Assuming that the observed transients are accreting neutron stars or black holes in X-ray binaries, we can estimate the mean accretion rate during an outburst, $\langle \dot{M}_{\mathrm{ob}} \rangle$, from the mean unabsorbed outburst flux. Following \citet{zand07}, we apply a correction factor of 3 to the mean 2-10 keV outburst luminosity (unabsorbed, inferred from spectral fitting) to obtain the 0.1-100 keV accretion luminosity $L_{\mathrm{acc}}$ (which is an approximation of the bolometric luminosity of the source). The mass-accretion rate during outburst is then estimated by employing the relation $\langle \dot{M}_{\mathrm{ob}} \rangle=R L_{\mathrm{acc}}/GM$, where $G=6.67 \times 10^{-8}~\mathrm{cm}^3~\mathrm{g}^{-1}~\mathrm{s}^{-2}$ is the gravitational constant. We adopt $M=1.4~\mathrm{M_{\odot}}$ and $R=10$~km for a neutron star accretor and $M=10~\mathrm{M_{\odot}}$ and $R=30$~km for the scenario of a black hole primary. 
Presuming that the observed outburst is typical, we convert the mass-accretion rate during outburst to a long-term averaged value, $\langle \dot{M}_{\mathrm{long}} \rangle$, by using the relation $\langle \dot{M}_{\mathrm{long}} \rangle=\langle \dot{M}_{\mathrm{ob}} \rangle \times t_{\mathrm{ob}} / t_{\mathrm{rec}}$, where $t_{\mathrm{ob}}$ is the outburst duration and $t_{\mathrm{rec}}$ is the system's recurrence time, i.e.,  the sum of the outburst and quiescence timescales. The factor $t_{\mathrm{ob}}/t_{\mathrm{rec}}$ represents the duty cycle of the system. 

The calculation of the time-averaged accretion rate, as described above, is subject to several uncertainties. Both the translation from the observed 2-10 keV luminosity to the bolometric luminosity, as well as the conversion to the mass-accretion rate are uncertain (the exact efficiency of converting gravitational potential energy to X-ray radiation is unknown). 
Furthermore, many X-ray transients show irregular outburst- and recurrence times, which makes it difficult to estimate their duty cycles and what we observe over the course of a few years may not be typical for their long-term accretion history. 
However, the quasi-daily \textit{Swift} monitoring observations of 2006-2007 provide an unique insight in the outburst behavior of these subluminous transients, allowing for a better estimate of their duty cycles than would be possible based on single, randomly spaced pointings alone. With the method described above, we can at least get an order of magnitude estimate of their time-averaged accretion rates.

An important caveat is that accretion flows around low luminosity (below a few percent of Eddington) black holes might be radiatively inefficient \citep[e.g.,][]{blandford99,narayan}. If this is the case, the mass-accretion rate as inferred from the X-ray luminosity can be severely underestimated. Thus, in particular the values inferred for the black hole scenario should be considered with caution (see Sec.~\ref{mdot_estimate}).

\begin{table*}
\begin{threeparttable}[tb]
\begin{center}
\caption{\textit{Chandra} positions (J2000) and errors of the transients that were found in archival  data.}
\begin{tabular}{l l l l l l}
\hline \hline
\ & \multicolumn{3}{c}{Coordinates\tnote{a}}  \\
\cline{2-4}
Name & R.A. & Decl. & Err. & Obs ID & Date \\
\ & (h m s) & $(^{\circ}~'~'')$ & ($''$) & \ & \  \\
\hline
AX J1745.6-2901 & 17:45:35.65 & $-$29:01:34.0 & 0.6 & 6639 & 2006-04-11\\
CXOGC J174535.5-290124 & 17:45:35.56 & $-$29:01:23.9 & 0.6 & 6644 & 2006-08-22 \\
CXOGC J174540.0-290005 & 17:45:40.06 & $-$29:00:05.5 & 0.6 & 6646 & 2006-10-29 \\
Swift J174553.7-290347 & 17:45:53.94 & $-$29:03:46.9 & 0.6 & 6363 & 2006-07-17 \\
Swift J174622.1-290634 & 17:46:22.25 & $-$29:06:32.5  & 1.3 & 6642 & 2006-07-04  \\
\hline
\end{tabular}
\label{tab:chandra}
\begin{tablenotes}
\item[a] The quoted position uncertainties ($1 \sigma$) were calculated by taking the square root of the quadric sum of the statistical error (from the \texttt{wavdetect} routine) and the uncertainty in absolute astrometry \citep[$0.6''$;][]{aldcroft00}.
\end{tablenotes}
\end{center}
\end{threeparttable}
\end{table*}

\section{X-ray lightcurves and spectra}\label{results}

\begin{figure*}[tb]
 \begin{center}
          \includegraphics[width=5.4cm]{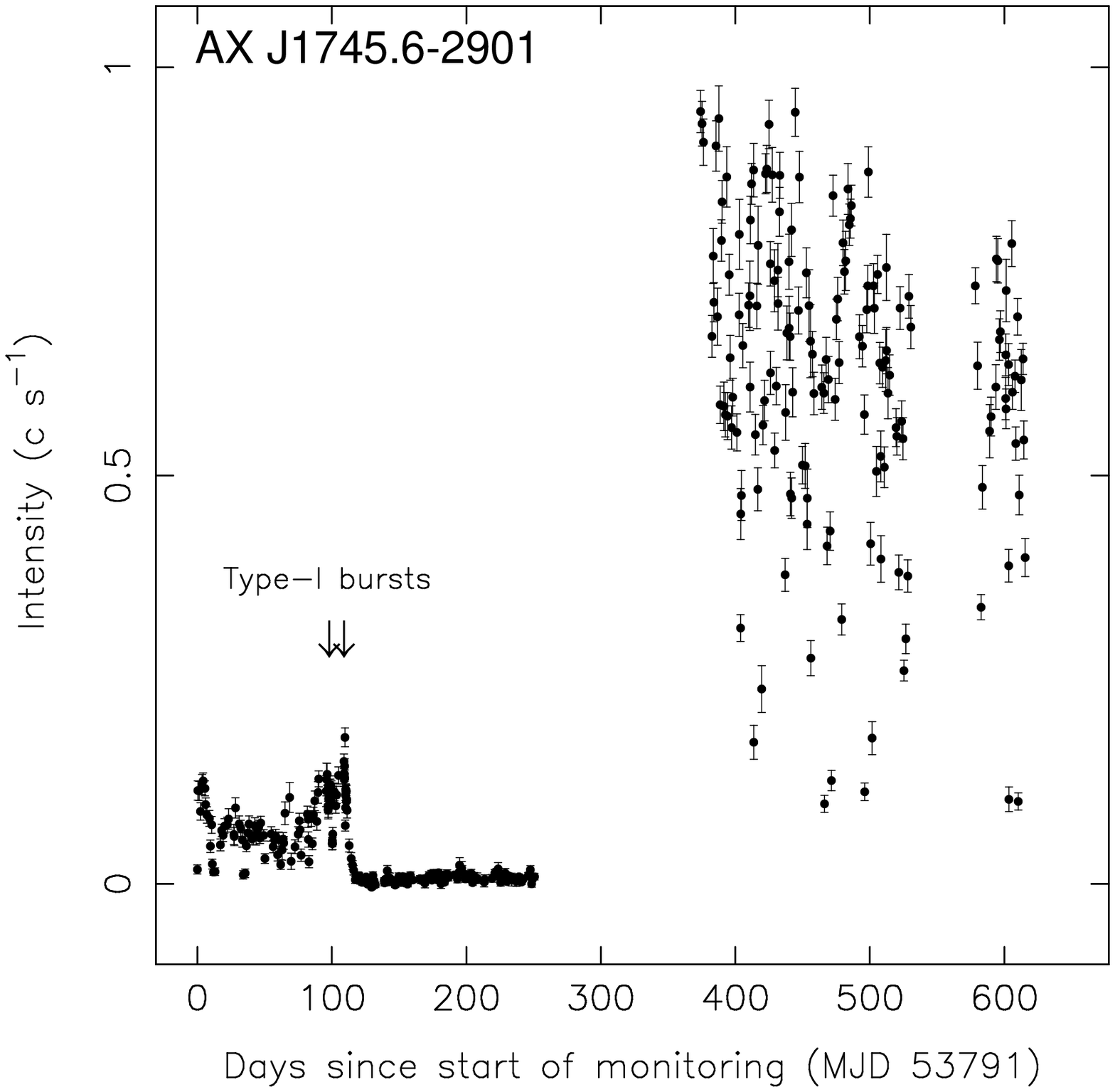}\hspace{0.5cm}\vspace{0.3cm}
      \includegraphics[width=5.4cm]{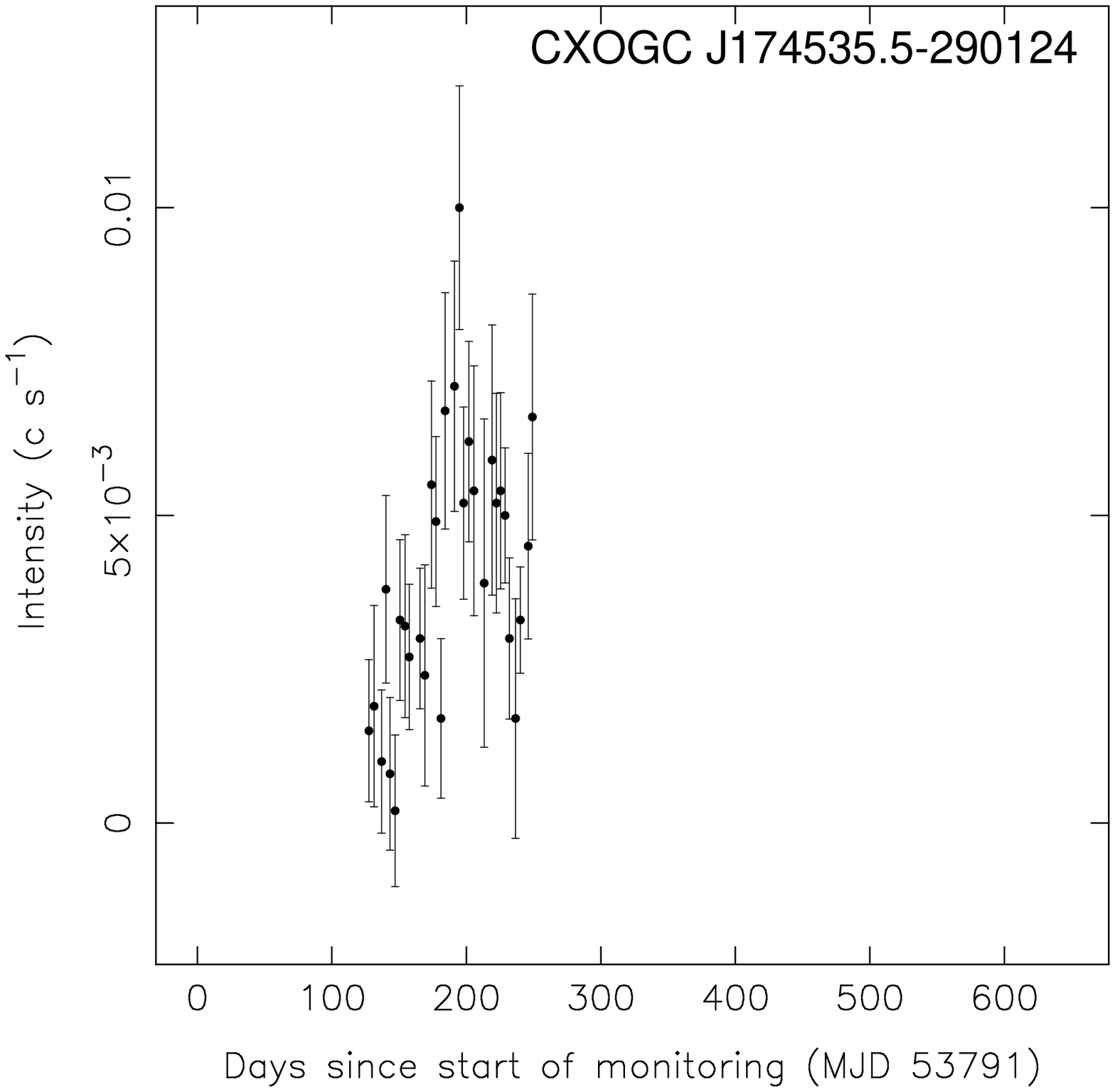}\hspace{0.5cm}
    \includegraphics[width=5.4cm]{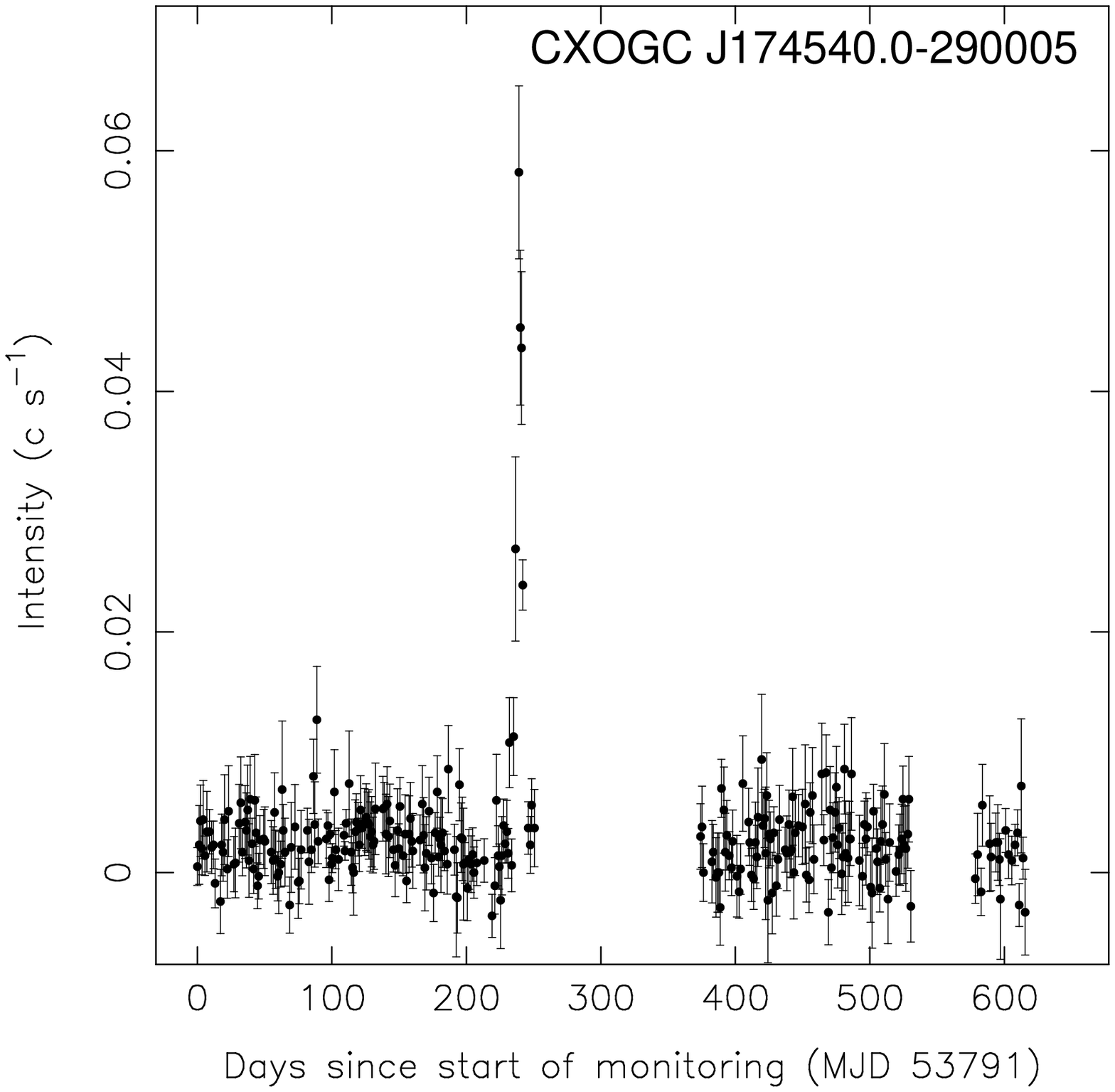}\hspace{0.5cm}
    \includegraphics[width=5.4cm]{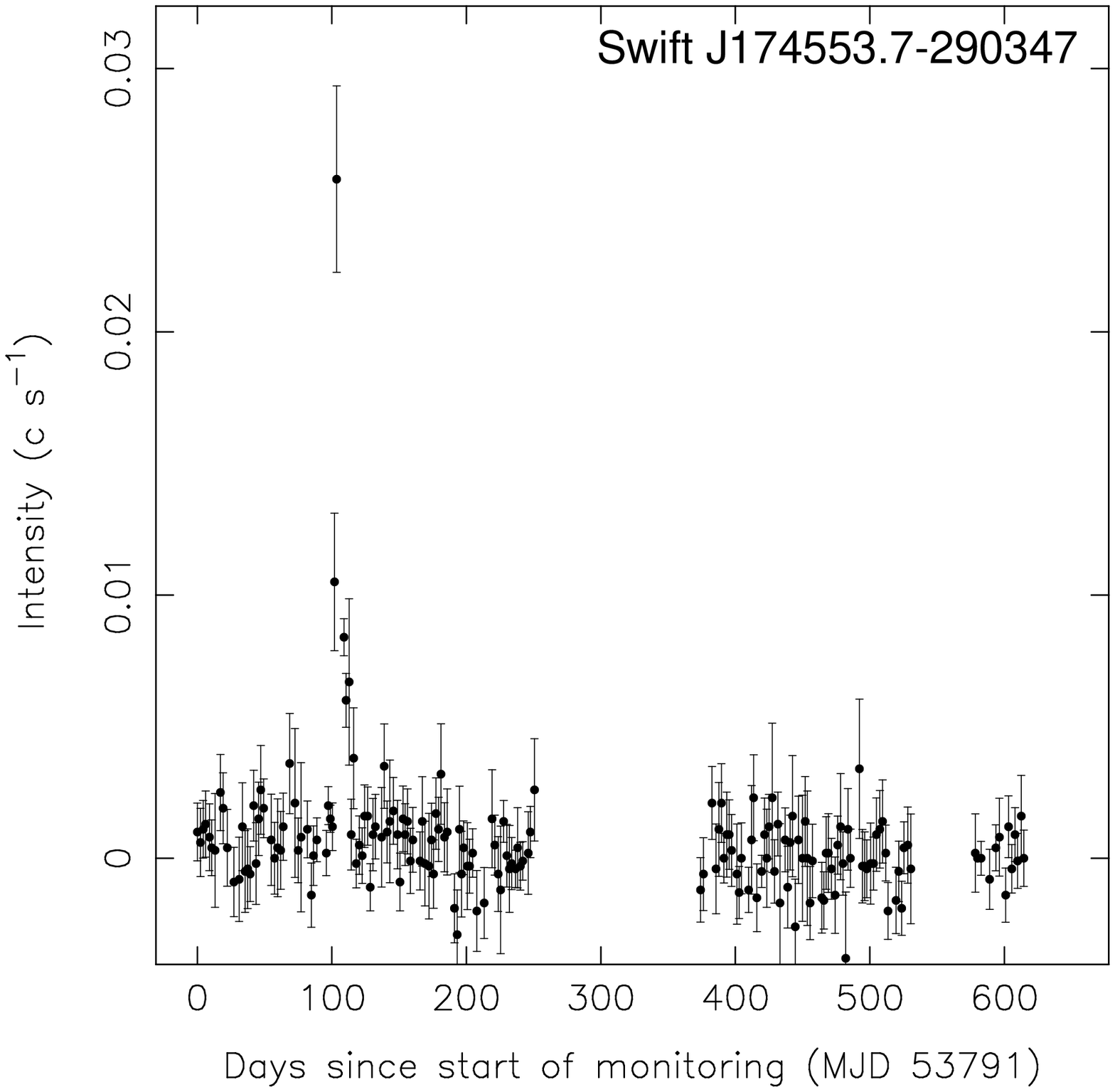}\hspace{0.5cm}\vspace{0.3cm}
      \includegraphics[width=5.4cm]{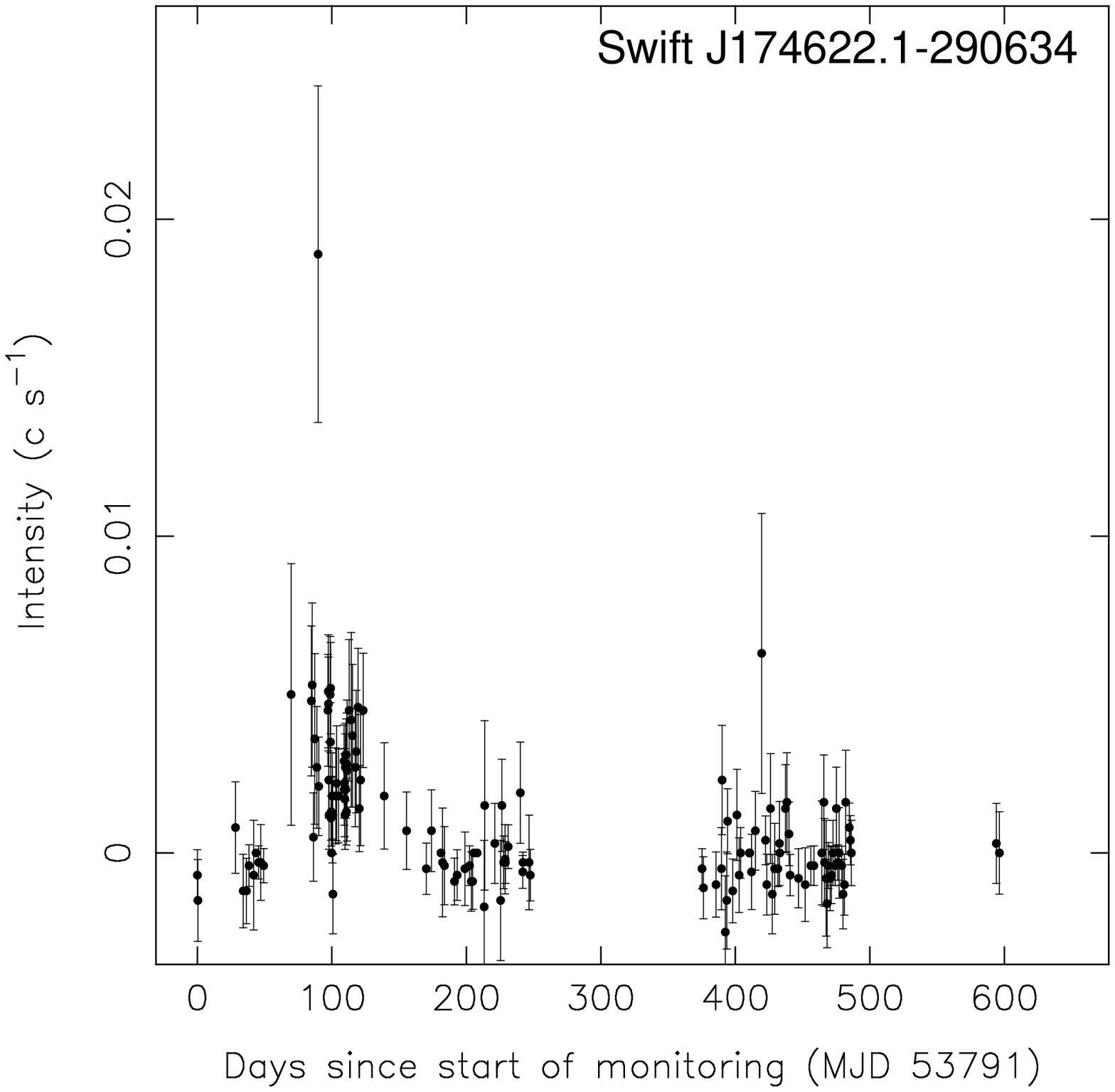}\hspace{0.5cm}
    \includegraphics[width=5.4cm]{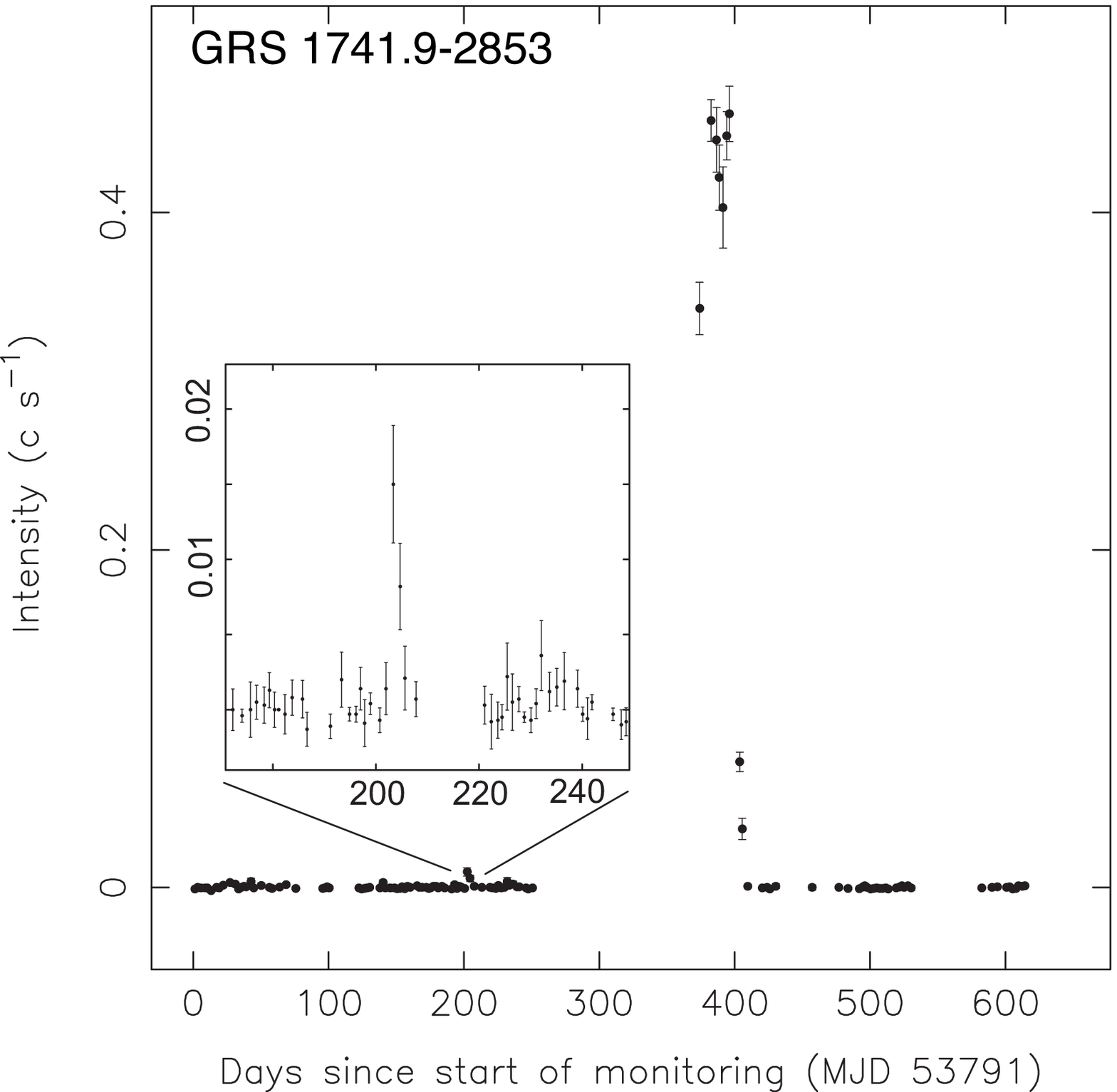}\hspace{0.5cm}
    \includegraphics[width=5.4cm]{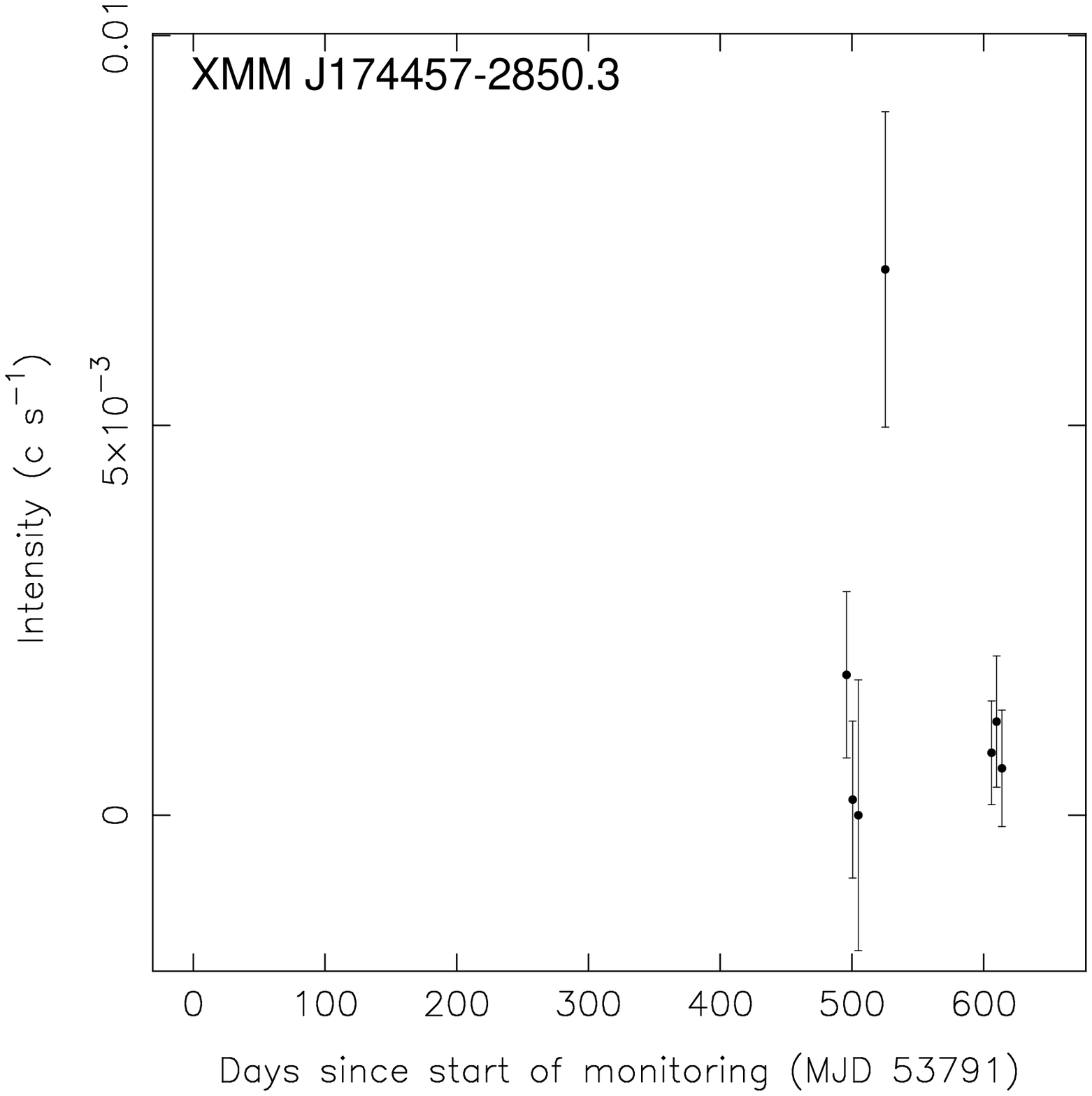}
    \end{center}
\caption[]{Background-corrected \textit{Swift}/XRT lightcurves of the seven transients that were found to be active during the \textit{Swift} monitoring campaign of the GC in 2006 and 2007. During days 252-374 after the start of the survey (February 24, 2006), no observations were carried out due to Solar constraints. In between days 533-579, all Swift's instruments were off-line due to a safe-hold event \citep{swift_offline07}. Furthermore, Swift J174622.1-290634, GRS 1741.9-2853 and XMM J174457-2850.3 were not always in FOV of the observations, so lightcurves of these three sources contain more data gaps than that of the other transients. We could only deduce information on the activity of CXOGC J174535.5-290124 from \textit{Swift} observations at times that the nearby, brighter transient AX J1745.6-2901 was not active. This was the case from late June till early November 2006, so the lightcurve of CXOGC J174535.5-290124 only covers this epoch. The plot of GRS 1741.9-2853 includes a zoom in of a small outburst that occurred in 2006. }
\label{fig:lc}
  \end{figure*}

\begin{figure*}[tb]
 \begin{center}
          \includegraphics[width=4.1cm, angle=270]{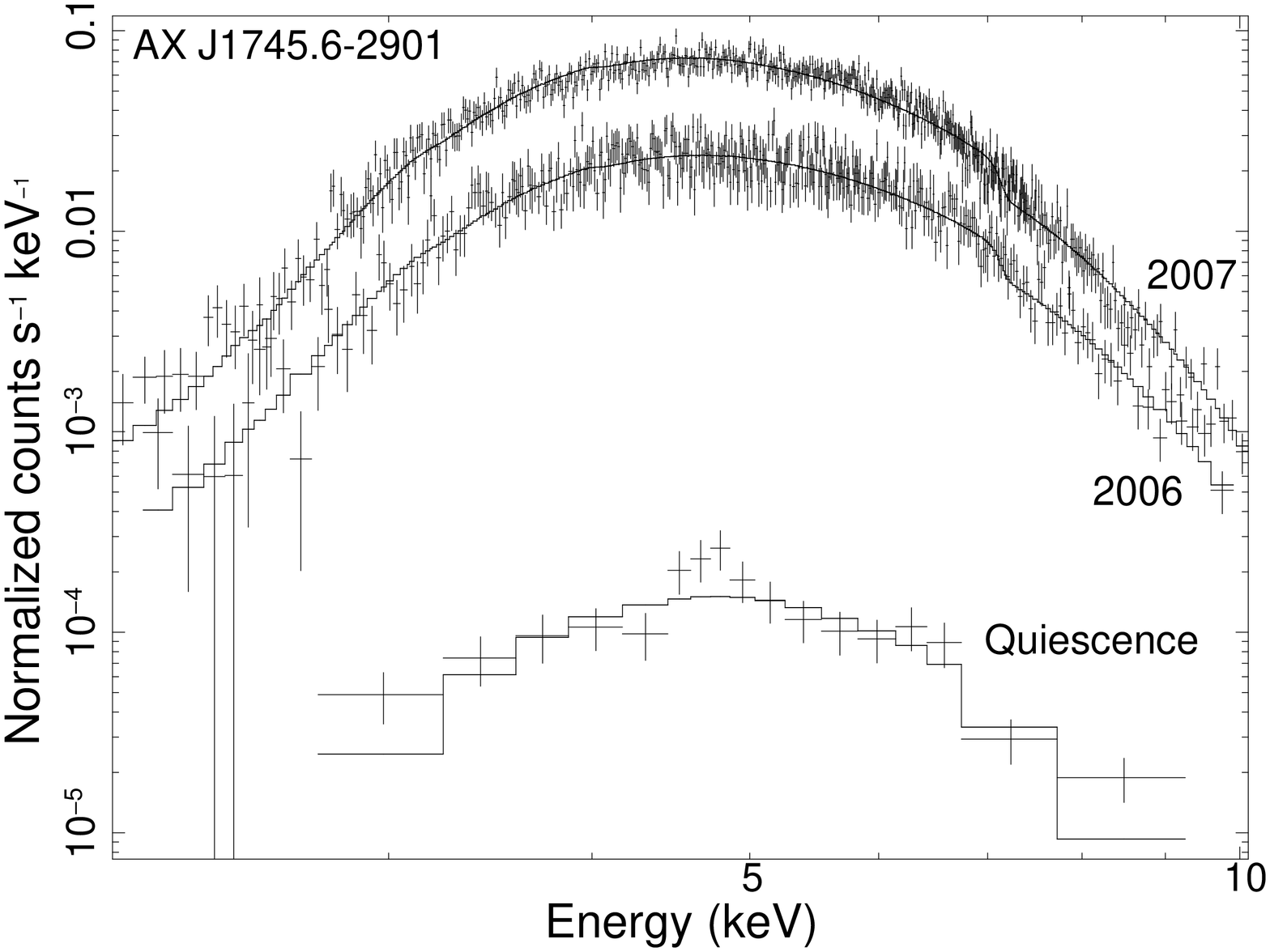}\hspace{0.5cm}\vspace{0.3cm}
      \includegraphics[width=4.1cm, angle=270]{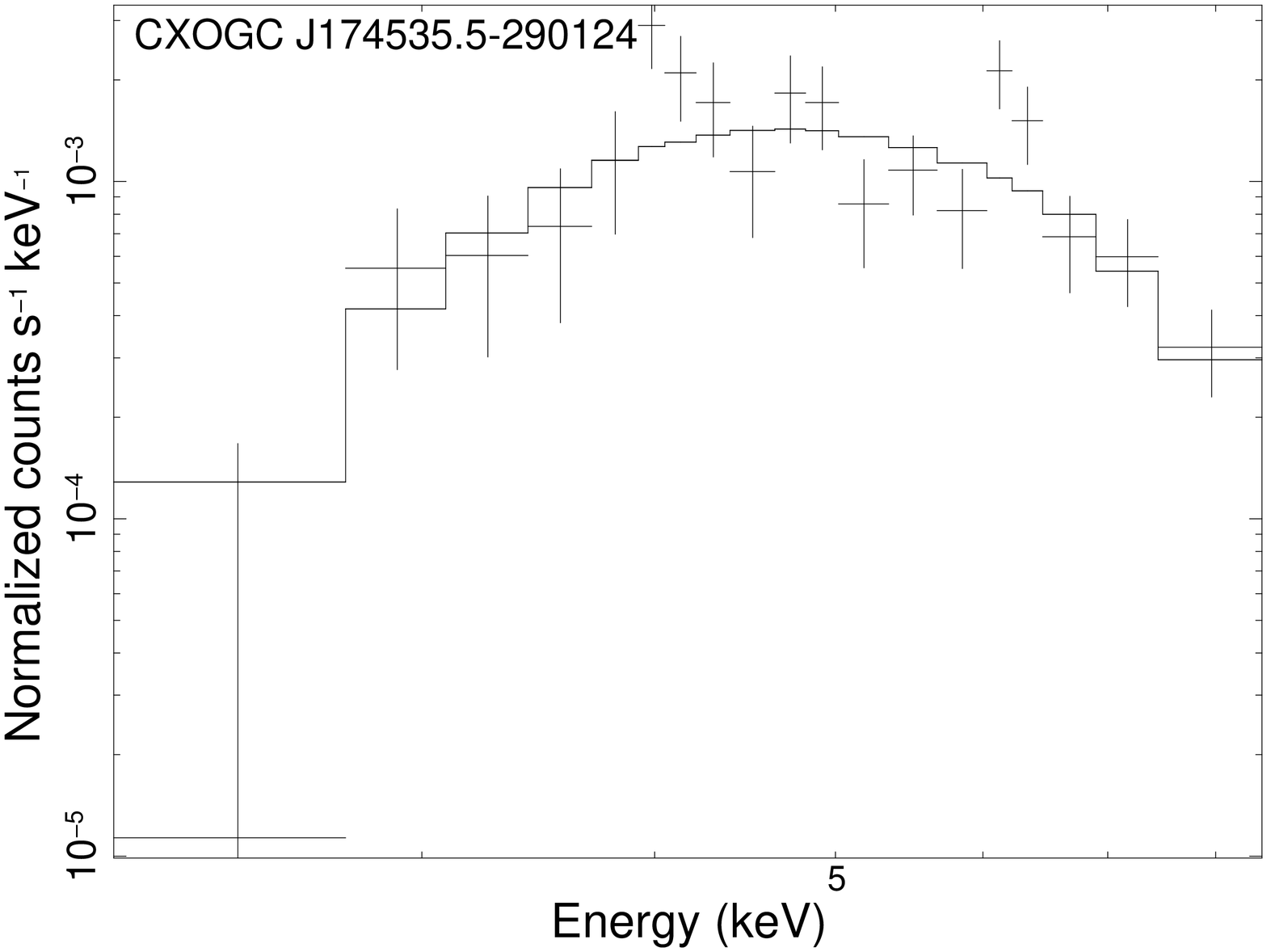}\hspace{0.5cm}
    \includegraphics[width=4.1cm, angle=270]{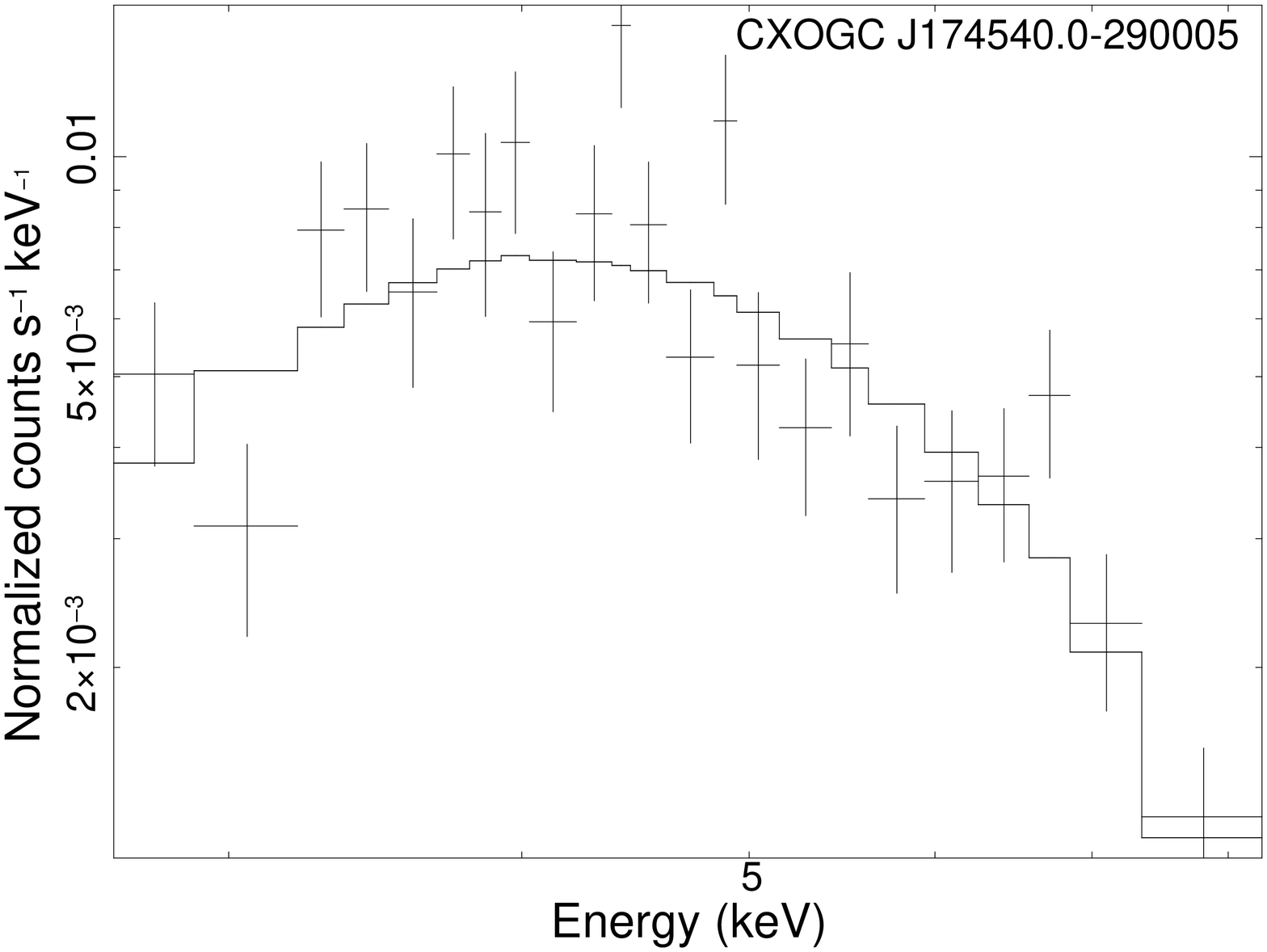}\hspace{0.5cm}
    \includegraphics[width=4.1cm, angle=270]{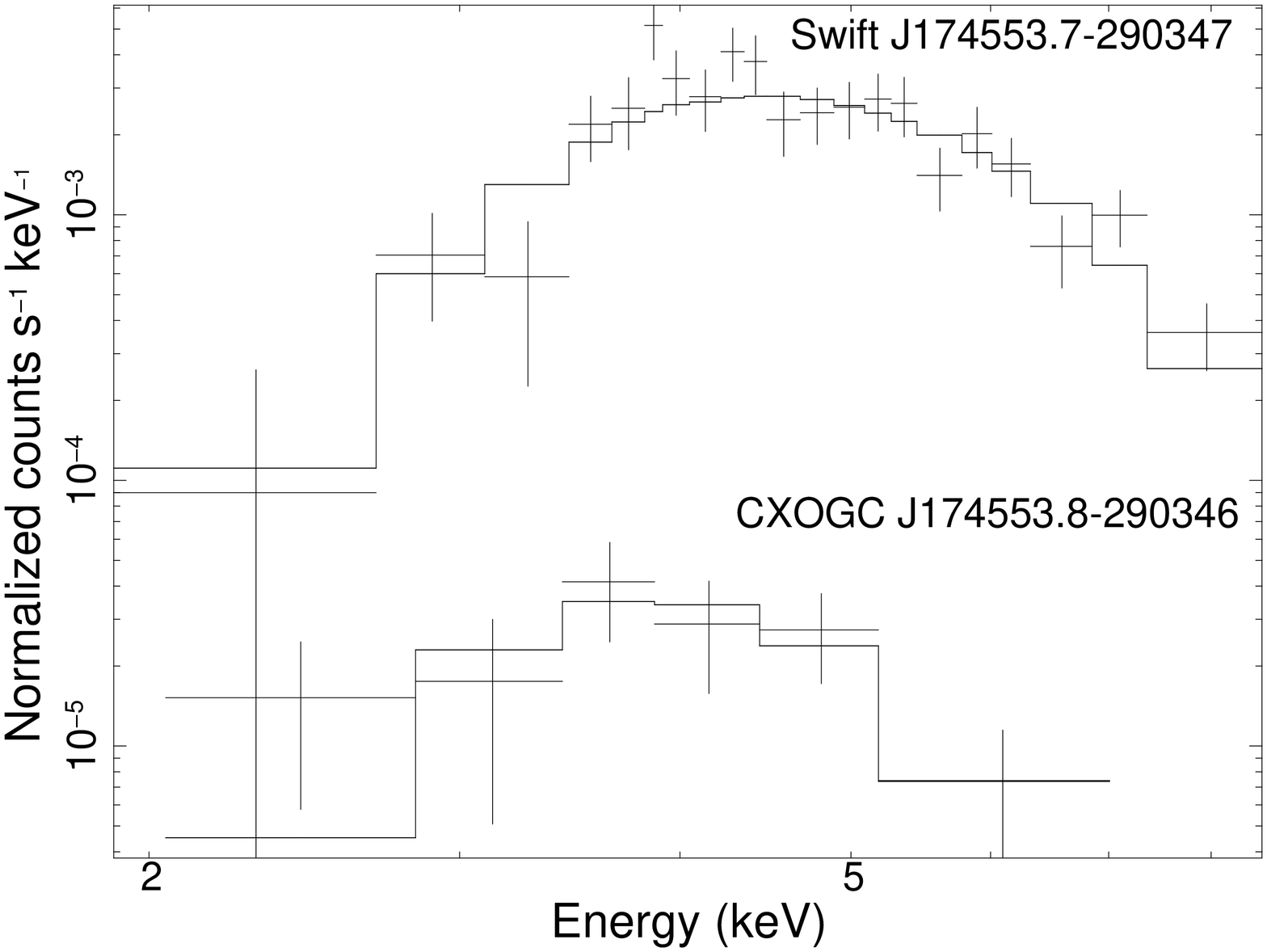}\hspace{0.5cm}\vspace{0.3cm}
      \includegraphics[width=4.1cm, angle=270]{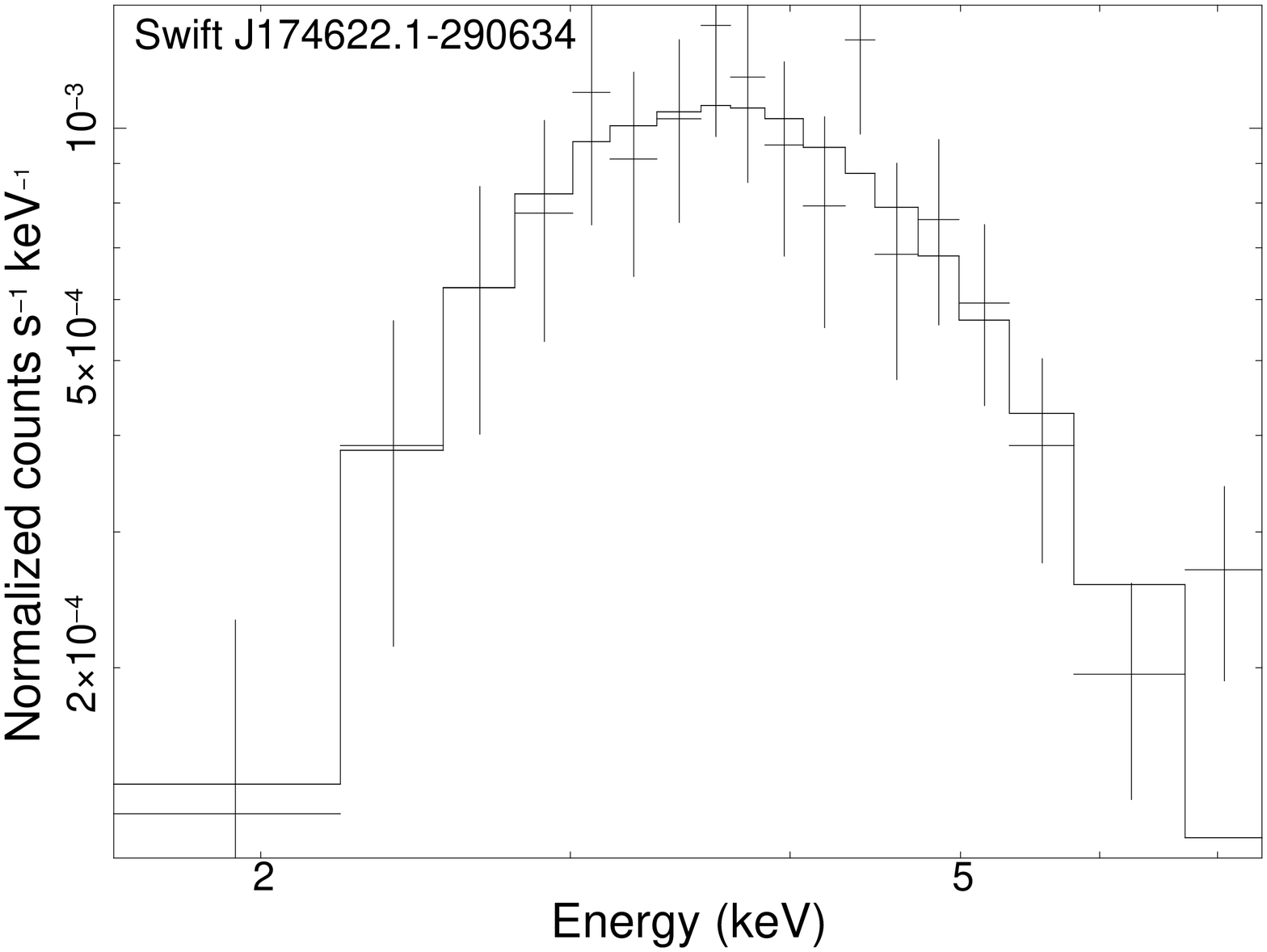}\hspace{0.5cm}
    \includegraphics[width=4.1cm, angle=270]{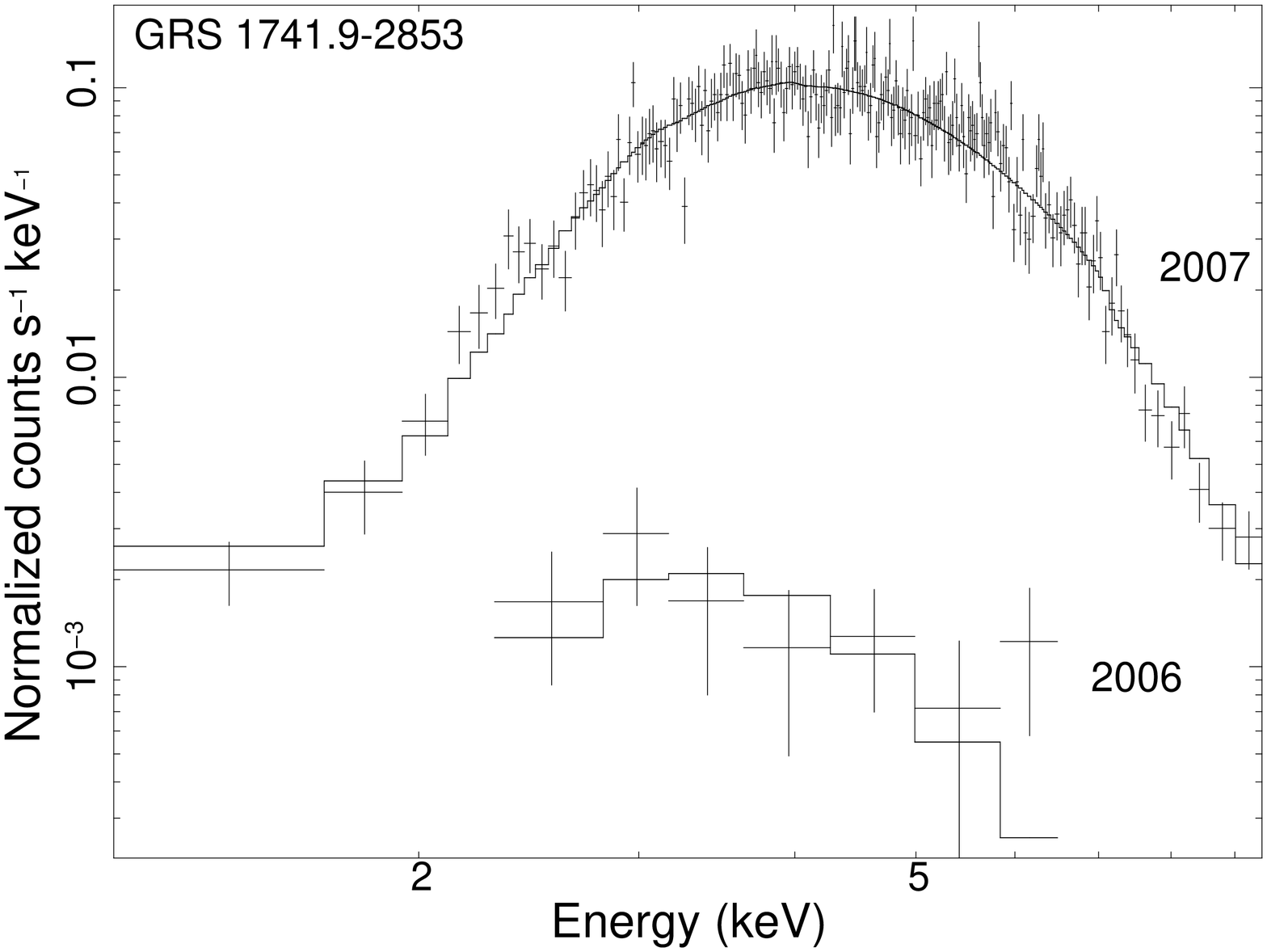}
    \end{center}
\caption[]{\textit{Swift}/XRT background-corrected spectra of the seven X-ray transients that were found to be active during the \textit{Swift} monitoring campaign of the field around $\mathrm{Sgr~A}^{\ast}$ in 2006 and 2007. The absorbed powerlaw model fit is plotted along with the data points for each source. For AX J1745.6-2901 three different spectra are shown: the average outburst spectrum of 2007 (upper), 2006 (middle) and the quiescent spectrum (lower) observed by \citet{muno04_apj613}. For the calculation of the average 2006 spectrum, we removed the two observations in which type-I X-ray bursts were detected. The plot of Swift J174553.7-290347 includes the X-ray spectrum of CXOGC J174553.8-290346 \citep[from][]{muno04_apj613}, which possibly represents the quiescent state of this transient. GRS 1741.9-2853 displayed separate outbursts in 2006 and 2007; average spectra of both outbursts are shown.}
\label{fig:spec}
  \end{figure*}

The background corrected lightcurves of the seven transients are displayed in Fig.~\ref{fig:lc} and their spectra are plotted in Fig.~\ref{fig:spec}. The X-ray properties of each individual source will be discussed below; a summary of the spectral parameters for all sources can be found in Table~\ref{tab:spectra}. All detected transients were heavily absorbed ($N_{H} \gtrsim 6 \times 10^{22}~\mathrm{cm}^{-2}$), consistent with what is observed for sources close to $\mathrm{Sgr~A}^{\ast}$. Therefore, throughout this paper we assume a distance of 8 kpc for all detected transients when calculating their 2-10 keV X-ray luminosities.

\begin{table*}
\begin{threeparttable}[t]
\begin{center}
\caption[]{{Overview of the spectral results for the seven active X-ray transients, obtained from the average outburst spectra.}}
\begin{tabular}{l l l l l l l l l l}
\hline
\hline
Source & Year & $N_{\mathrm{H}}$\tnote{a} & $\Gamma$ & red. $\chi^{2}$ & $F_{\mathrm{X, abs}}$\tnote{b} & $F_{\mathrm{X, unabs}}$\tnote{c} & $F_{\mathrm{X, peak}}$\tnote{d} & $L_{\mathrm{X}}\tnote{e}$ & $L_{\mathrm{X, peak}}\tnote{f}$\\
\hline 
AX J1745.6-2901 & 2006 & $23.1 \pm 1.3$ & $2.3 \pm 0.2$ & 1.11 & $14.7^{+0.3}_{-0.2} $ & $50.4^{+4.6}_{-3.8}$ & $120$ & $39$ & $92$ \\
 & 2007 & $24.9^{+0.6}_{-0.7}$ & $2.8 \pm 0.1$ & 1.14 & $ 44.8^{+0.4}_{-0.3}$ & $205^{+9}_{-11}$ & $800$ & $160$ & $610$ \\
CXOGC J174553.5-290124 & 2006 & $14.2^{+5.1}_{-7.4}$ & $1.1^{+1.2}_{-1.1}$ & 1.40 & $ 1.25^{+0.25}_{-0.21}$ & $2.25 ^{+1.10}_{-0.43}$ & $4.0$ & $1.7$ & $3.0$ \\
CXOGC J174540.0-290005 & 2006 & $8.63^{+5.66}_{-5.00}$ & $1.4^{+1.0}_{-0.9}$ & 1.54 & $ 7.81^{+0.92}_{-0.88}$ &  $ 12.5^{+4.9}_{-2.2}$ & $30$ & $9.6$ & $23$ \\
Swift J174553.7-290347 & 2006 & $24.4^{+10.1}_{-7.7}$ & $3.0^{+1.6}_{-1.1}$ & 1.23 & $ 1.53^{+0.22}_{-0.24}$ & $7.73^{+18.0}_{-3.49}$ & $26$ & $5.9$ & $20$ \\
Swift J174622.1-290634 & 2006 & $11.7^{+6.1}_{-3.9}$ & $3.3^{+1.4}_{-1.0}$ & 0.57 & $ 0.468\pm0.064$ & $1.55^{+2.06}_{-0.55}$ & $9.1$ & $1.2$ & $7.0$ \\
GRS 1741.9-2853 & 2006 & $14$ fix & $5.0^{+2.5}_{-2.7}$ & 0.84 & $0.646^{+0.424}_{-0.199}$ & $5.07^{+4.36}_{-2.38}$ & $12$ & $3.9$ & $9.2$ \\
 & 2007 & $14.0^{+1.0}_{-0.9}$ & $2.6 \pm 0.2$ & 1.15 & $61.6\pm1.2$ & $175^{+17}_{-14}$ & $260$ & $130$ & $200$ \\
XMM J174457-2850.3\tnote{g} & 2007 & $6$ fix& $1.3$ fix& & $0.21$ & $0.29$ & $1.4$ & $0.22$ & $1.1$  \\
\hline
\end{tabular}
\label{tab:spectra}
\begin{tablenotes}
\item[a]Hydrogen column density in units of $10^{22}~\mathrm{H~cm}^{-2}$.
\item[b]Mean 2-10 keV absorbed outburst flux in units of $10^{-12}~\mathrm{erg~cm}^{-2}~\mathrm{s}^{-1}$.
\item[c]Mean 2-10 keV unabsorbed outburst flux in units of $10^{-12}~\mathrm{erg~cm}^{-2}~\mathrm{s}^{-1}$.
\item[d]Unabsorbed peak flux observed during the outburst in units of $10^{-12}~\mathrm{erg~cm}^{-2}~\mathrm{s}^{-1}$.
\item[e]The mean outburst X-ray luminosity, in units of $10^{34}~\mathrm{erg~s}^{-1}$, is calculated from the mean unabsorbed flux by adopting a distance of 8 kpc for all sources.
\item[f]The peak X-ray luminosity, in units of $10^{34}~\mathrm{erg~s}^{-1}$, is calculated from the peak unabsorbed flux by adopting a distance of 8 kpc for all sources.
\item[g]Fluxes for XMM J174457-2850.3 were deduced using PIMMS, with $N_{\mathrm{H}}$ and $\Gamma$ fixed at the values obtained by \citet{sakano05}.
\end{tablenotes}
\end{center}
\end{threeparttable}
\end{table*}

\subsection{AX J1745.6-2901}\label{subsec:bron1}
The start of the \textit{Swift}/GC monitoring observations in 2006 immediately revealed the new X-ray transient Swift J174535.5-290135, which is located $\sim 1.5'$ SE from $\mathrm{Sgr~A}^{\ast}$ \citep{kennea06_atel753}. This X-ray source remained active for approximately 16 weeks until it returned to quiescence in late June, 2006. Renewed activity of the system was reported in February 2007 \citep{wijnands07_atel1006, kuulkers07_atel1005}, and the \textit{Swift}/GC monitoring observations suggest that it remained as such for more than a year, as it was active until the campaign ended in November 2007 (see Fig.~\ref{fig:lc}). We note that the monitoring campaign continued in 2008 and that the source was detected throughout 2008. However, a detailed discussion of those observations are beyond the scope of our paper. The detection of eclipses with an 8.4 hours period seen in \textit{XMM-Newton} observations \citep{porquet07}, positively identify Swift J174535.5-290135 with the \textit{ASCA} detected eclipsing X-ray burster AX J1745.6-290 \citep{maeda1996}. 
In addition, the \textit{Chandra} position of AX J1745.6-290 (see Table~\ref{tab:chandra}) is consistent with that of the X-ray source CXOGC J174535.6-290133 \citep{muno03}, which likely represents the quiescent counterpart of the system. 

Figure~\ref{fig:lc} displays the activity of AX J1745.6-2901 during the 2006-2007 \textit{Swift} campaign. In 2006, the outburst reached a peak luminosity of $9.2 \times 10^{35}~\mathrm{erg~s}^{-1}$, while the average outburst luminosity was $3.9 \times 10^{35}~\mathrm{erg~s}^{-1}$ (both in the 2-10 keV energy band). For an outburst duration of at least 16 weeks (AX J1745.6-2901 might have been active before the start of the \textit{Swift} monitoring campaign), we can deduce a fluency of $\gtrsim 1.8 \times 10^{-4}~\mathrm{erg~cm}^{-2}$. In 2007, the system was active again, but with an higher average luminosity of $1.6 \times 10^{36}~\mathrm{erg~s}^{-1}$ and a reached peak value of $6.1 \times 10^{36}~\mathrm{erg~s}^{-1}$ (both 2-10 keV).  Different outburst luminosities have been reported for AX J1745.6-2901 in the past; in October 1993, the source was detected at a luminosity of $2 \times 10^{35}~\mathrm{erg~s}^{-1}$, while in October 1994 it became as bright as $9 \times 10^{35}~\mathrm{erg~s}^{-1}$ \citep[both values are in the 3-10 keV band,][]{maeda1996}. 

Before and after the 6-week epoch in 2007 that the \textit{Swift} observatory was offline due to a safe-hold event \citep[][this corresponds to days 533-579 in the lightcurves displayed in Fig.~\ref{fig:lc}]{swift_offline07}, AX J1745.6-2901 was active at similar count rates.
We have inspected proprietary \textit{XMM-Newton} data of the GC performed on September 6, 2007 (Degenaar et al. in preparation), i.e., halfway the interval that the \textit{Swift} observatory was offline. AX J1745.6-2901 was clearly detected during that observation, which demonstrates that the source remained active all through the 2007 \textit{Swift} monitoring campaign. For an outburst duration of 34 weeks, the 2-10 keV fluency of the 2007 outburst is then $5.7 \times 10^{-3}~\mathrm{erg~cm}^{-2}$. However, this inferred value should be considered a lower limit, since we also found AX J1745.6-2901 to be active during all \textit{Swift}/GC monitoring observations in 2008, at a flux similar to that of 2007 (the source was also reported active during \textit{Chandra} observations carried out in 2008, see \citet{heinke08} and \citet{deeg08_atel_gc}). This suggests that the outburst observed in 2007, continued in 2008 and thus has a duration of at least 1.5 years. For that outburst length, the fluency increases to $6.5 \times 10^{-3}~\mathrm{erg~cm}^{-2}$ (2-10 keV), and will become even larger if the outburst continues.

Between 1999 and 2002, the GC was observed several times with \textit{Chandra} \citep{muno03,muno04_apj613}. Thus, if the observed long outburst duration of AX J1745.6-2901 is typical, the source likely resided in quiescence for at least 4 years. However, the quiescent timescale must be less than 13 years, the time since the \textit{ASCA} discovery \citep{maeda1996}.  
Estimating the long-term time-averaged mass-accretion rate for AX J1745.6-2901 is difficult due to the different outburst durations and luminosities the system displays. To get a rough estimate, we will assume that an outburst duration of 1.5 years and a 2-10 keV outburst luminosity of $2 \times 10^{36}~\mathrm{erg~s}^{-1}$ are typical for the source. The duty cycle of this neutron star system then ranges from $10\%$ for $t_{\mathrm{q}}\sim 13$~yr up to $30\%$ for $t_{\mathrm{rec}} \sim 4$~yr. This results in an estimated long-term time-averaged accretion rate of $\sim (5-15) \times 10^{-11}~\mathrm{M_{\odot}~yr}^{-1}$ (see Table~\ref{tab:mdot}). This value might be a lower limit, since AX J1745.6-2901 possibly exhibited more outbursts like the smaller one observed in 2006. On the other hand, the observed long outburst of 1.5 years might not be typical for the system, in which case this estimate would be an upper limit on the time-averaged mass-accretion rate.

To compare the outburst spectrum of AX J1745.6-2901 with the likely quiescent counterpart of the source (CXOGC J174535.6-290133), we downloaded the reduced data of the \textit{Chandra} monitoring campaign that are made available online\footnote{Available at http://www.astro.psu.edu/users/niel/galcen-xray-data/galcen-xray-data.html \citep{muno04_apj613}.\label{foot:chan}}. After the spectrum was grouped to contain at least 20 photons per bin, we fitted it with an absorbed powerlaw model with the hydrogen column density fixed at the 2006 outburst value ($N_{H}=23.1 \times 10^{22}~\mathrm{cm}^{-2}$). This resulted in a powerlaw index of $\Gamma=1.8 \pm 0.5$ and an unabsorbed 2-10 keV flux of $7.7^{+0.4}_{-0.3} \times 10^{-14}~\mathrm{erg~cm}^{-2}~\mathrm{s}^{-1}$. The inferred 2-10 keV luminosity is $5.9 \times 10^{32}~\mathrm{erg~s}^{-1}$. The quiescent spectrum is plotted in Fig.~\ref{fig:spec} along with the average outburst spectra of 2006 and 2007.\\

\noindent
 \textbf{Type-I X-ray bursts}\\
The \textit{Swift}/GC monitoring observations detected two type-I X-ray bursts from AX J1745.6-2901. The times at which these bursts occurred are indicated in Fig.~\ref{fig:lc}.
The first burst was observed on June 3, 2006, and had an exponential decay timescale of $\sim 10$~s (see Fig.~4a). Due to the sudden increase in count rate associated with the X-ray burst, the XRT instrument automatically switched from PC to WT mode. There is no burst data available during this switch, which took about 3 seconds. We extracted the spectrum of the first 3 seconds of the observed burst peak and fitted it to an absorbed blackbody model with the hydrogen column density fixed at $\mathrm{N_{H}}=23.1 \times 10^{22}~\mathrm{cm}^{-2}$, the value inferred from the mean outburst spectrum of 2006. This yielded $kT=1.7^{+1.9}_{-0.6}~\mathrm{keV}$ and a radiating surface area of $10^{+12}_{-6}~ \mathrm{km}$ (assuming $d=8~$kpc). The 0.01-100 keV peak flux inferred from our spectral fit is $1.3^{+2.1}_{-0.1} \times 10^{-8}~\mathrm{erg~cm}^{-2}~\mathrm{s}^{-1}$ (corrected for absorption), which translates into an observed peak luminosity of $9.6 \times 10^{37}\mathrm{~erg~s}^{-1}$. However, the true burst peak was likely missed due to the automatic switch of XRT modes. If we extrapolate the burst lightcurve to the time $t=-3$~s (i.e., the time of the mode-switch), we can deduce a 0.01-100 keV peak luminosity of $1.3 \times 10^{38}\mathrm{~erg~s}^{-1}$. Although the true peak of the type-I X-ray bursts will remain uncertain, it is clear that it likely was close to the Eddington luminosity for a neutron star \citep[$2.0 \times 10^{38}\mathrm{~erg~s}^{-1}$ for a hydrogen-rich and $3.8 \times 10^{38}\mathrm{~erg~s}^{-1}$ for a hydrogen-poor photosphere; e.g.,][]{kuulkers03_xrb}.

Another burst was observed on June 14, 2006, which had an exponential decay timescale of $\sim$20~s (see Fig.~4b). This time, no automated switch of XRT modes occurred, so that the burst was fully detected in the PC mode. Due to the high count rate associated with the burst, the PC image was severely piled-up and a proper spectral fitting of the burst peak was not possible. Therefore, we used the burst count rate to find the peak flux and luminosity. To obtain the correct count rates, the observed ones have to be corrected for the loss in photons caused by bad columns
and pixels using an exposure map, and a pile-up correction needs to be applied. For the latter, we extracted the source photons from an annular source region, avoiding the piled-up inner pixels. We determined the proper correction factor for the observed PC count rate following analysis threads on the \textit{Swift} webpages. This way, we found that the burst must have reached a peak count rate of $15~\mathrm{cnts~s}^{-1}$ in the PC mode. Employing PIMMS with a hydrogen column density of $N_{\mathrm{H}}=23.1 \times 10^{22}~\mathrm{cm}^{-2}$ (the 2006 outburst value) and temperatures of $kT=1.0-3.0~\mathrm{keV}$ (roughly the range inferred for the first burst), we can estimate an unabsorbed 0.01-100 keV flux of $(0.68-1.0) \times 10^{-8}~\mathrm{erg~cm}^{-2}~\mathrm{s}^{-1}$. The corresponding 0.01-100 keV peak luminosity is $(0.52-1.1)\times 10^{38}\mathrm{~erg~s}^{-1}$, i.e., comparable to the X-ray burst that occurred on June 3, 2006.

\begin{figure*}[tb]
 \begin{center}
 \subfigure[]{\label{fig:bursts-a}\includegraphics[width=5.2cm]{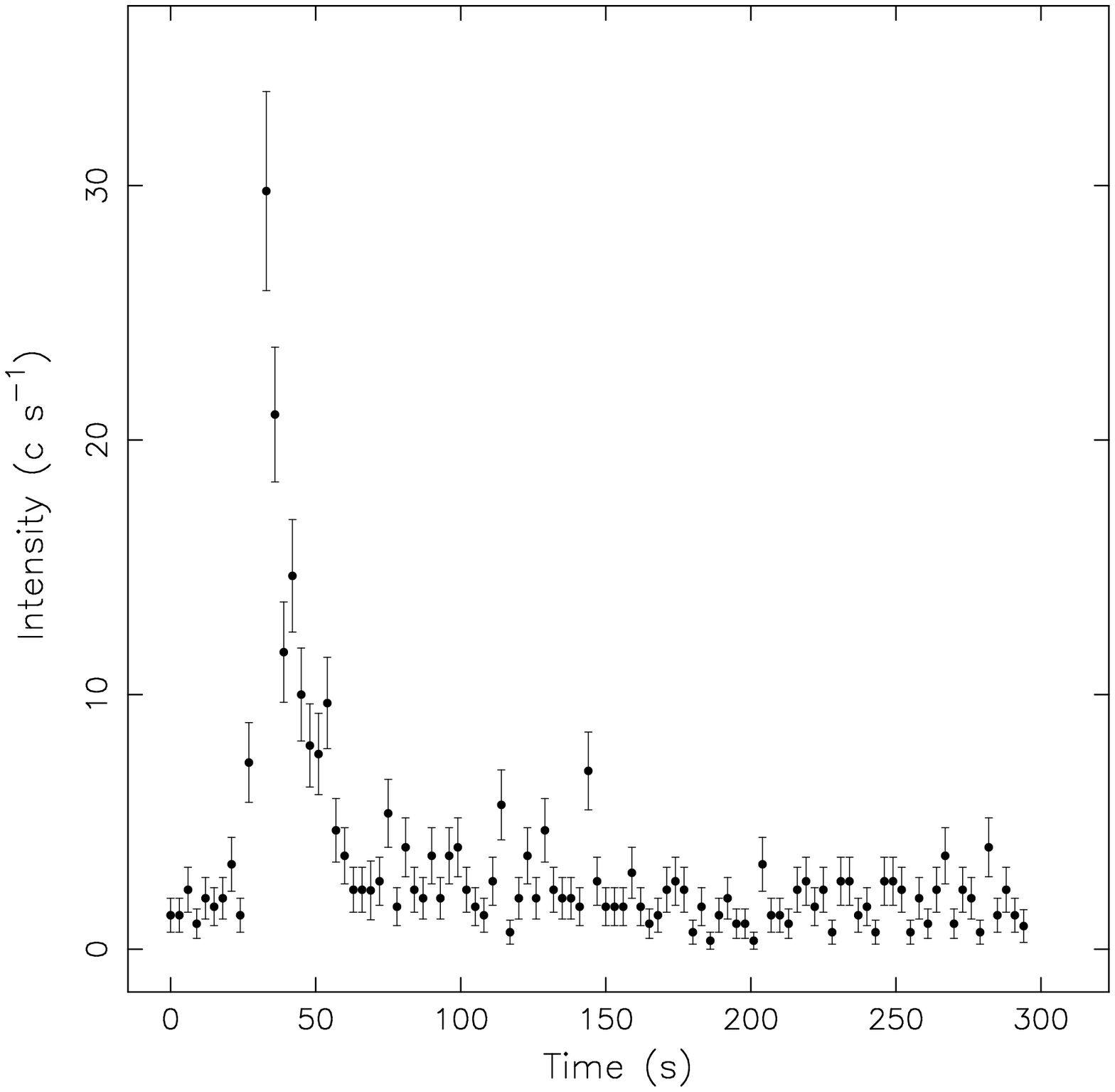}}\hspace{0.5cm}
 \subfigure[]{\label{fig:bursts-b}\includegraphics[width=5.2cm]{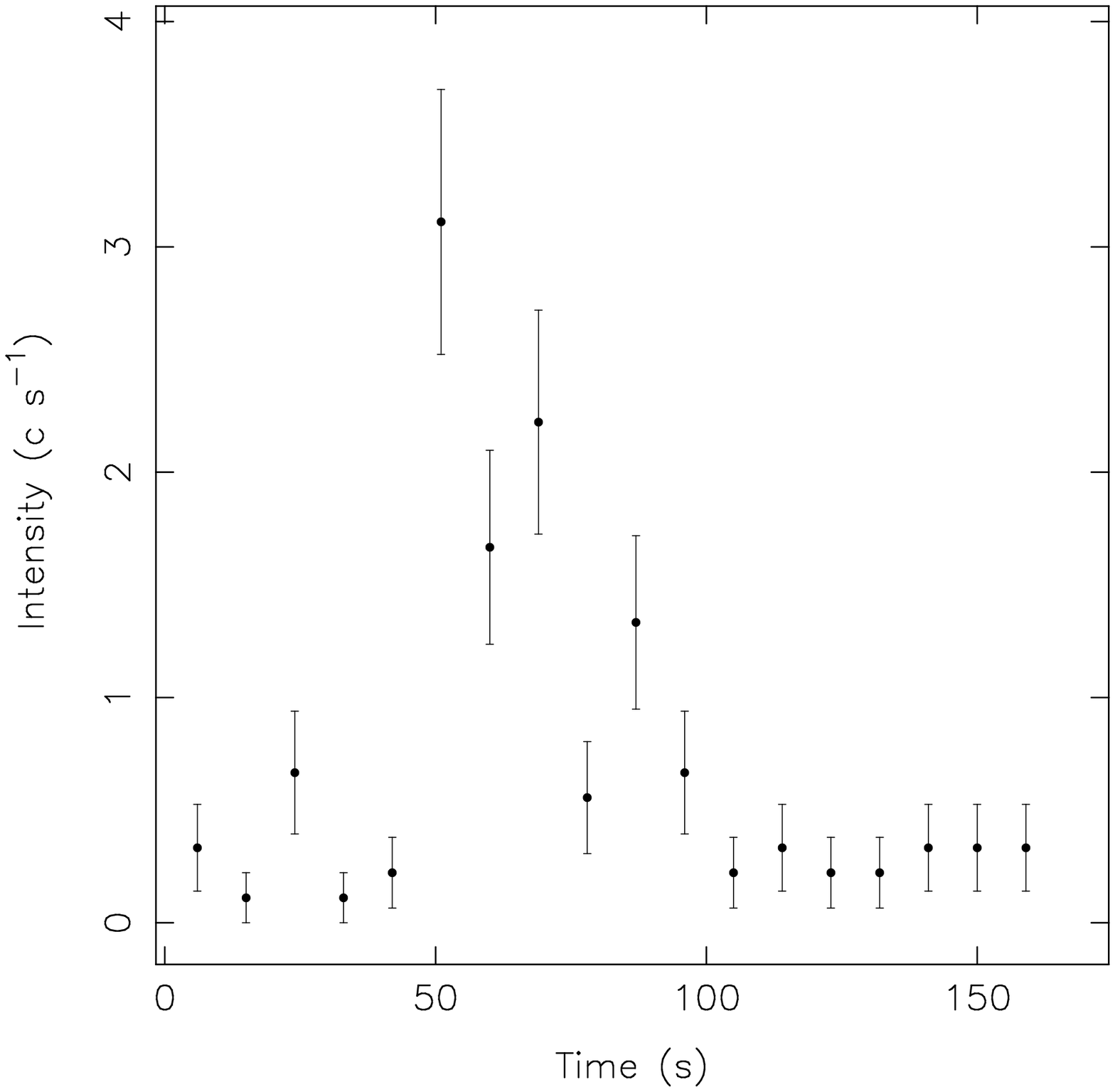}}
    \end{center}
\caption[]{a) Lightcurve of the type-I X-ray burst from AX J1745.6-2901 that occurred on June 3, 2006, and was observed in both PC and WT mode. b) A second burst from this source was observed on June 14, 2006, which was solely detected in PC mode.}
  \end{figure*}

\subsection{CXOGC J174535.5-290124}
By the beginning of August 2006, a transient was detected in outburst approximately $1.3'$ SE from $\mathrm{Sgr~A}^{\ast}$. The XRT position for this source, which is listed in Table~\ref{tab:vfxts}, is only $\sim 14''$ away from the above discussed AX J1745.6-2901, which had returned to quiescence a month earlier. We obtained an improved position for this transient from an archival \textit{Chandra} observation performed on August 22, 2006 (see Table~\ref{tab:chandra}) and find that its coordinates are not consistent with the \textit{Chandra} position of AX J1745.6-2901 (Table~\ref{tab:chandra}), but do coincide with that of the known X-ray transient CXOGC J174535.5-290124 \citep[][]{muno05_apj622}. 

CXOGC J174535.5-290124 is a subluminous X-ray transient that was discovered during a \textit{Chandra} campaign of the GC \citep{muno04_apj613}. Whereas the source was not detected in 1999 and 2000 \citep[yielding a 2-8 keV upper limit for the quiescent luminosity of $L_{\mathrm{X}}<9 \times 10^{30}~\mathrm{erg~s}^{-1}$;][]{muno05_apj622}, it was found in outburst with \textit{Chandra} on several occasions between 2001 and 2005, displaying typical 2-8 keV luminosities of $10^{33-34}~\mathrm{erg~s}^{-1}$  \citep{muno05_apj622, wijn05_atel638, deeg08_atel_gc}. The source was also detected in outburst during \textit{XMM-Newton} observations obtained in September 2006, when it displayed a 2-10 keV X-ray luminosity of $2 \times 10^{34}~\mathrm{erg~s}^{-1}$ \citep{wijn06_atel892}. This is in agreement with the 2006 \textit{Swift} data of CXOGC J174535.5-290124 (see Table~\ref{tab:spectra}), which showed an average outburst luminosity of $1.7 \times 10^{34}~\mathrm{erg~s}^{-1}$ and an observed peak luminosity of $3.0 \times 10^{34}~\mathrm{erg~s}^{-1}$ (both in the 2-10 keV energy band). The source was observed in outburst until the \textit{Swift} observations stopped in November 2006. The outburst of late 2006 thus had a duration of at least 12 weeks. This yields a lower limit on the outburst fluency of $1.6 \times 10^{-5}~\mathrm{erg~cm}^{-2}$ in the 2-10 keV energy band.

The nearby transient AX J1745.6-2901 is typically a factor 10-100 brighter during outburst than CXOGC J174535.5-290124, and due to their small separation, we cannot deduce any information on CXOGC J174535.5-290124 from \textit{Swift} data when AX J1745.6-2901 is active. However, \textit{Chandra} does have the required spatial resolution to separate these two transients, even when AX J1745.6-2901 is in outburst. Inspection of archival \textit{Chandra} data of both 2006 and 2007 revealed that CXOGC J174535.5-290124 was in outburst simultaneously with AX J1745.6-2901 in April 2006 (Obs ID 6639), although it was not active during \textit{Chandra} observations carried out in May, June and early July 2006. Thus, CXOGC J174535.5-290124 must have returned to quiescence by the end of April 2006, but it reappeared in August 2006, when \textit{Swift} detected the source. Since AX J1745.6-2901 was continuously active during the \textit{Swift}/GC monitoring observations of 2007, we cannot deduce any information on the activity CXOGC J174535.5-290124 from the 2007 \textit{Swift} data. The source was not found active in archival \textit{Chandra} data obtained in February and July 2007, nor in proprietary \textit{Chandra} observations carried out in March, April and May 2007 (Degenaar et al. 2008 in preparation).
In 2008, CXOGC J174535.5-290124 is reported active during pointed \textit{Chandra}/HRC-I observations performed on May 10-11 \citep{deeg08_atel_gc}. During that observation the 2-10 keV X-ray luminosity was approximately $2 \times 10^{33}~\mathrm{erg~s}^{-1}$, i.e., a factor of 10 lower than outburst level detected with \textit{Swift} in 2006. 

Despite its low peak luminosity, CXOGC J174535.5-290124 appears to be active quite regularly. However, its duty cycle is not completely clear; the \textit{Swift} observations show that in 2006 the system was in quiescence for about 3 months in between two outbursts, while \textit{Chandra} data suggests that it was likely quiescent for more than 6 months in 2007 (if an outburst duration of $\sim 3$~months is typical). Tentatively assuming a recurrence time of 3-12 months and a typical outburst duration of 12 weeks, the duty cycle for CXOGC J174535.5-290124 is $\sim 20-50 \%$. The various detections of CXOGC J174535.5-290124 vary between a few times $10^{33-34}~\mathrm{erg~s}^{-1}$, so we adopt a mean 2-10 keV outburst luminosity of $1 \times 10^{34}~\mathrm{erg~s}^{-1}$. This results in a long-term averaged accretion rate of $ 5 \times 10^{-13}~\mathrm{M}_{\odot}~\mathrm{yr}^{-1} \lesssim \langle \dot{M}_{\mathrm{long}} \rangle \lesssim 1 \times 10^{-12}~\mathrm{M}_{\odot}~\mathrm{yr}^{-1}$ for a neutron star primary or $ 7 \times 10^{-14}~\mathrm{M}_{\odot}~\mathrm{yr}^{-1} \lesssim \langle \dot{M}_{\mathrm{long}} \rangle \lesssim 2 \times 10^{-13}~\mathrm{M}_{\odot}~\mathrm{yr}^{-1}$ in case of a black hole accretor (see Table~\ref{tab:mdot}).

\subsection{CXOGC J174540.0-290005}
In late October 2006, \citet{kennea06_atel920} reported on activity from an X-ray transient,
Swift J174540.2-290005, located $\sim 20''$ N from $\mathrm{Sgr~A}^{\ast}$. In an archival \textit{Chandra} observation performed on August 22, 2006, we find one X-ray source within the XRT error radius of Swift J174540.2-290005 (see Table~\ref{tab:vfxts}). The \textit{Chandra} position of this source (see Table~\ref{tab:chandra}) is consistent with that of CXOGC J174540.0-290005 \citep{muno05_apj622}, positively identifying Swift J174540.2-290005 with this \textit{Chandra}-discovered X-ray transient. CXOGC J174540.0-290005 was detected in outburst only once before, in 2003, when it displayed a luminosity of $3.4 \times 10^{34}~\mathrm{erg~s}^{-1}$ \citep[2-8 keV,][]{muno05_apj622}. This is a factor of a few lower than the peak luminosity of $2.3 \times 10^{35}~\mathrm{erg~s}^{-1}$ that was detected by \textit{Swift} in 2006 (2-10 keV, see Table~\ref{tab:spectra}). \citet{muno05_apj622} derived an upper limit for the quiescent luminosity of this system of $<4 \times 10^{31}~\mathrm{erg~s}^{-1}$ (2-8~keV). 

The 2006 outburst of CXOGC J174540.0-290005 lasted almost 2 weeks, and the inferred outburst fluency is $1.3 \times 10^{-5}~\mathrm{erg~cm}^{-2}$ (2-10 keV).  No other outburst from CXOGC J174540.0-290005 was detected during the \textit{Swift} observing campaign of 2006 and 2007. If the observed outburst duration of 2 weeks is typical for this source, than its outbursts are most easily missed. However, CXOGC J174540.0-290005 was in FOV during the entire \textit{Swift} campaign of 2006, which encompassed almost daily observations and lasted for 35 weeks. No activity from the source was detected in 33 weeks prior to the outburst that occurred in late October. Therefore, we can assume an upper limit on the duty cycle of $\lesssim 6\%$. This is consistent with the fact that no outburst was detected during the 2007 \textit{Swift} monitoring observations. However, since the source was detected with \textit{Chandra} in 2003 \citep{muno05_apj622}, we can also assume that the recurrence time of the system is less than 3 years. From this we can deduct that the duty cycle is likely more than $1\%$. Using these two bounds combined with an averaged 2-10 keV outburst luminosity of $1 \times 10^{35}~\mathrm{erg~s}^{-1}$, we can put a limit on the time-averaged mass-accretion rate of $ 3 \times 10^{-13}~\mathrm{M}_{\odot}~\mathrm{yr}^{-1} \lesssim \langle \dot{M}_{\mathrm{long}} \rangle \lesssim 1.5 \times 10^{-12}~\mathrm{M}_{\odot}~\mathrm{yr}^{-1}$ for a neutron star compact primary, or $ 4 \times 10^{-14}~\mathrm{M}_{\odot}~\mathrm{yr}^{-1} \lesssim \langle \dot{M}_{\mathrm{long}} \rangle \lesssim 2.1 \times 10^{-13}~\mathrm{M}_{\odot}~\mathrm{yr}^{-1}$ in case it is a black hole.

Following the reported activity of Swift J174540.2-290005 \citep{kennea06_atel920}, \citet{wang06_atel935} performed IR observations of the source field on October 30-31, 2006. Whereas no sources within the XRT source position uncertainty showed an expected increase in IR brightness \citep[e.g.,][]{clark00_IR,russell06}, \citet{wang06_atel935} concluded that none of them was the counterpart to CXOGC J174540.0-290005. However, we note that the \textit{Swift}/XRT data shows that at the time of the reported IR observations, the X-ray outburst had already ceased and any correlated IR luminosity might have returned to its pre-outburst level accordingly.

\subsection{Swift J174553.7-290347}\label{subsec:bron4}
A fourth X-ray transient, which we designate Swift J174553.7-290347 (see Table~\ref{tab:vfxts}), is located $\sim 4.5'$ SW from $\mathrm{Sgr~A}^{\ast}$ and was found active for a duration of approximately 2 weeks in June 2006 (see Fig.~\ref{fig:lc}). The source reached a peak luminosity of $ 2.0 \times 10^{35}~\mathrm{erg~s}^{-1}$, while the average outburst luminosity was $ 4.9 \times 10^{34}~\mathrm{erg~s}^{-1}$ (both 2-10 keV). We were able to obtain an improved position of Swift J174553.7-290347 from an archival \textit{Chandra} observation carried out on June 17, 2006 (see Table~\ref{tab:chandra}). The source coordinates suggests a possible association with the \textit{Chandra} detected X-ray point source CXOGC J174553.8-290346 \citep{muno03}, although the offset between the source positions is $\sim 1''$. During the \textit{Chandra} campaign of the GC \citep[][]{muno03,muno04_apj613,muno05_apj622}, CXOGC J174553.8-290346 was detected as a low luminosity X-ray source ($L_{\mathrm{X}} \sim10^{32}~\mathrm{erg~s}^{-1}$, 2-8 keV) that showed no signs of long- or short-term variability \citep{muno03}. 

The spectral shape of CXOGC J174553.8-290346 is not reported in literature, but the reduced \textit{Chandra} data (both source and background spectra as well as proper response files) from the campaign are made available online (see footnote~\ref{foot:chan} on page~\pageref{foot:chan}). For comparison with the current outburst data, we downloaded the reduced \textit{Chandra} data and fitted the background corrected spectrum with an absorbed powerlaw model (after the spectra were grouped to contain at least 20 photons per bin). With the absorption column density fixed at the outburst value of Swift J174553.7-290347, $N_{H}=24.4~\times 10^{22}~\mathrm{cm}^{-2}$, this results in a fit with an unusual steep spectrum; $\Gamma=5.5\pm 2.0$. The 2-10 keV unabsorbed X-ray flux for this fit is $5.0^{+6.7}_{-3.0} \times 10^{-14}~\mathrm{erg~cm}^{-2}~\mathrm{s}^{-1}$ and the associated X-ray luminosity would be $ 3.8 \times 10^{32}~\mathrm{erg~s}^{-1}$. Leaving the hydrogen column density as a free parameter results in a fit with $N_{H}=11.6^{+13.2}_{-7.0}~\times 10^{22}~\mathrm{cm}^{-2}$ and $\Gamma =3.1^{+1.5}_{-1.9}$ and an X-ray luminosity of $ \sim 6.8 \times 10^{31}~\mathrm{erg~s}^{-1}$ (2-10 keV). Both the \textit{Swift} outburst spectrum of Swift J174553.7-290347 and the \textit{Chandra} spectrum of CXOGC J174553.8-290346 are plotted in combination with the fitted models in Fig.~\ref{fig:spec} (the plotted spectral model for CXOGC J174553.8-290346 is for $N_{H}=24.4~\times 10^{22}~\mathrm{cm}^{-2}$). 

During the entire 2006-2007 \textit{Swift} campaign, the new X-ray transient Swift J174553.7-290347 only displayed this 2-week outburst (see Fig.~\ref{fig:lc}), for which we can infer a 2-10 keV outburst fluency of $ 8.0 \times 10^{-6}~\mathrm{erg~cm}^{-2}$. The source was not detected during 22 weeks of consecutive observations in 2007, which we can use as a lower limit on the quiescent timescale of this source (which is consistent with the 2006 behavior). Thus, the duty cycle of Swift J174553.7-290347 is likely less than $8\%$. The estimate for the long-term average accretion rate of this transient is then $\lesssim 10^{-12}~\mathrm{M}_{\odot}~\mathrm{yr}^{-1}$ for a neutron star X-ray binary or $\lesssim 2 \times 10^{-13}~\mathrm{M}_{\odot}~\mathrm{yr}^{-1}$ in case of an accreting black hole.

\subsection{Swift J174622.1-290634}
Approximately $11'$ SW from $\mathrm{Sgr~A}^{\ast}$, the X-ray transient Swift J174622.1-290634 is active from mid-May till late-June 2006 (see Fig.~\ref{fig:lc}). We obtained improved coordinates for this new X-ray transient from an archival \textit{Chandra} observation carried out on July 7, 2006, during which the source was detected (see Table~\ref{tab:chandra}). This system cannot be identified with any known X-ray source \citep[it was outside FOV of the \textit{Chandra} monitoring campaign of the GC;][]{muno03, muno04_apj613,muno05_apj622}. The average outburst luminosity during the \textit{Swift}/XRT observations was $1.2 \times 10^{34}~\mathrm{erg~s}^{-1}$ and the observed peak luminosity was $7.0 \times 10^{34}~\mathrm{erg~s}^{-1}$ (both in the 2-10 keV energy band). The 2-10 keV outburst fluency for the 5-week outburst of Swift J174622.1-290634 is $5.0 \times 10^{-6}~\mathrm{erg~cm}^{-2}$.

Swift J174622.1-290634 lies relatively far from $\mathrm{Sgr~A}^{\ast}$ and was not always within the FOV, due to varying pointing centers and roll-angles of the \textit{Swift}/XRT observations. However, the observations were spread such that if the observed outburst duration of 5 weeks is typical for the source, any other outburst occurring during the 2006 monitoring campaign would have been detected by \textit{Swift}/XRT. During the 6-week interval that the \textit{Swift} observatory was offline in 2007, Swift J174622.1-290634 could in principle have experienced an accretion outburst of 5 weeks. However, the system was not detected during \textit{XMM-Newton} observations of the GC performed on September 6, 2007 (i.e., halfway the interval that the \textit{Swift} observatory was offline), indicating that this is not the case. We therefore assume that the source was in quiescence for the entire 2007 \textit{Swift} monitoring campaign, which lasted for 31 weeks. The duty cycle of Swift J174622.1-290634 is thus likely less than $14\%$, which makes its time-averaged accretion rate $\lesssim 4 \times 10^{-13} ~\mathrm{M}_{\odot}~\mathrm{yr}^{-1}$ for an accreting neutron star or $\lesssim 6 \times 10^{-14} ~\mathrm{M}_{\odot}~\mathrm{yr}^{-1}$ for a black hole X-ray binary.\\

\subsection{GRS 1741.9-2853}
The neutron star X-ray transient GRS 1741.9-2853 (located $\sim 10 '$ NE from $\mathrm{Sgr~A}^{\ast}$) was in FOV during most of the \textit{Swift} monitoring observations (see Fig.~\ref{fig:lc}).  The source has been detected in an active state several times since its initial discovery in 1990 \citep{sunyaev1990}, displaying typical peak luminosities of a few times $10^{36}~\mathrm{erg~s}^{-1}$  \citep[e.g.,][]{muno03_grs, wijn06_monit}. In September 2006, GRS 1741.9-2853 displayed some low level activity, lasting approximately a week (see Fig.~\ref{fig:lc}, $\sim 200$ days after the start of the monitoring observations). The source reached a peak luminosity of $8.9 \times 10^{34}~\mathrm{erg~s}^{-1}$ (2-10 keV), which is an order of magnitude lower than its full outburst luminosity, but still about 1000 times higher than its quiescent level \citep[$\sim 10^{32}~\mathrm{erg~s}^{-1}$ in the 2-8 keV band;][]{muno03_grs}. The fluency of this small outburst is $2.6 \times 10^{-6}~\mathrm{erg~cm}^{-2}$ (2-10 keV). 

Renewed activity from GRS 1741.9-2853 was reported in early 2007, as observed with \textit{INTEGRAL} \citep{kuulkers07_atel1008}, \textit{Swift} \citep{wijnands07_atel1006}, \textit{XMM-Newton} \citep{porquet07} and \textit{Chandra} \citep{muno07_atel1013}. During its 2007 activity, three type-I X-ray bursts were reported \citep{wijnands07_atel1006, porquet07} and several of such thermonuclear bursts have been observed in the past \citep[see][and references therein]{muno03_grs}. 
GRS 1741.9-2853 was seen active right from the start of the 2007 \textit{Swift}/GC on March 3, 2007. It remained as such for approximately 5 weeks, displaying an average 2-10 keV outburst luminosity of $1.3 \times 10^{36}~\mathrm{erg~s}^{-1}$, until it returned to quiescence by the beginning of April. During this outburst, \textit{Swift}/XRT detected a peak luminosity of $2.0 \times 10^{36}~\mathrm{erg~s}^{-1}$ (2-10 keV). For the observed outburst duration of 5 weeks, the 2-10 keV fluency of the 2007 outburst is $5.3 \times 10^{-4}~\mathrm{erg~cm}^{-2}$. 

However, GRS 1741.9-2853 was already seen active during \textit{INTEGRAL} observations performed on February 15, i.e., 2 weeks before the start of the \textit{Swift}/GC campaign. Moreover, \citet{wijnands07_atel1006} noted that GRS 1741.9-2853  is located within the $3'$ error circle of an X-ray burst detected by the Burst Alert Telescope (BAT) onboard \textit{Swift} on January 22, 2007 \citep{grb_grs07}. As there were no other sources detected within the BAT error circle, it is likely that GRS 1741.9-2853 was the origin of this burst, suggesting that the source was already active for over 8 weeks before the start of the \textit{Swift}/GC campaign. Therefore, the outburst fluency inferred from the \textit{Swift}/XRT observations should be considered as a lower limit and the true value might be $>1.4 \times 10^{-3}~\mathrm{erg~cm}^{-2}$ (2-10 keV), in case the outburst lasted 13 weeks, or longer.

Small outbursts like the one occurring in 2006 with $L_{\mathrm{X}}^{\mathrm{peak}}\sim 10^{35}~\mathrm{erg~s}^{-1}$ and $t_{\mathrm{ob}}=1$~week, have a negligible effect on the total mass-accretion rate, when compared to longer and brighter outbursts like the one observed in 2007. Therefore, we will not include the 2006 outburst in calculating the mass-accretion rate for GRS 1741.9-2853 and assume a minimal quiescent timescale of 35 weeks (the span of the 2006 monitoring observations). Adapting a typical outburst duration of 13 weeks (which is likely the minimum duration of the 2007 outburst), we can than place an upper limit on the the duty cycle of GRS 1741.9-2853 of $\lesssim 30 \%$. 
On the other hand, GRS 1741.9-2853 has been detected at 2-10 keV X-ray luminosities of $\sim 10^{36}~\mathrm{erg~s}^{-1}$ for a total of 5 times since its initial discovery 18 years ago \citep[see][for the long-term lightcurve of this source, showing its various outbursts from 1990 till 2005]{wijn06_monit}. Therefore, we assume a lower limit on the duty cycle of $\gtrsim 7\%$. Combining these bounds with a typical 2-10 keV outburst luminosity of $\sim 10^{36}~\mathrm{erg~s}^{-1}$, we estimate a long-term accretion rate of $ 2 \times 10^{-11}~\mathrm{M}_{\odot}~\mathrm{yr}^{-1} \lesssim \langle \dot{M}_{\mathrm{long}} \rangle \lesssim 8 \times 10^{-11}~\mathrm{M}_{\odot}~\mathrm{yr}^{-1}$ for GRS 1741.9-2853.

\subsection{XMM J174457-2850.3}\label{subsec:bron7}
XMM J174457-2850.3 is an X-ray transient located about $13.7\arcmin$ NE from $\mathrm{Sgr~A}^{\ast}$. The source was discovered in 2001 \citep{sakano05}, using XMM-Newton observations, at a peak luminosity of $ 5 \times 10^{34}~\mathrm{erg~s}^{-1}$, but with a quiescent luminosity of $1.2 \times 10^{32}~\mathrm{erg~s}^{-1}$ (both in the 2-10 keV energy range). Since then, the source has been repeatedly reported active at luminosities ranging from a few times $10^{33}~\mathrm{erg~s}^{-1}$ up to $\sim 10^{36}~\mathrm{erg~s}^{-1}$ \citep{wijn06_monit,muno07_atel1013}. 

Due to its large angular separation from  $\mathrm{Sgr~A}^{\ast}$, there were only 16 pointings (spaced between July and November 2007) during the \textit{Swift}/GC campaign, in which XMM J174457-2850.3 was in FOV (see Fig.~\ref{fig:lc}). Restricted by a small number of photons, we could not fit to the spectrum of XMM J174457-2850.3. We therefore employed PIMMS to convert the observed XRT count rates to fluxes using an absorbed powerlaw model with $N_{\mathrm{H}}=6.0 \times 10^{22}~\mathrm{cm}^{-2}$ and $\Gamma=1.3$ \citep[as found by][]{sakano05}. During the first set of 6 observations (performed between July 5 and July 14, 2007; for a total of 6 ksec), the source had a 2-10 keV X-ray luminosity of $\sim 1.5 \times 10^{33}~\mathrm{erg~s}^{-1}$. This is at a similar level as was found for the source in February 2007 by \citet{muno07_atel1013}. However, on August 4, the source was clearly detected during a single $\sim 1.7~$ksec observation at $L_{\mathrm{X}} \sim 1.1 \times 10^{34}~\mathrm{erg~s}^{-1}$ (2-10 keV). XMM J174457-2850.3 was again in FOV during a series of 6 \textit{Swift} monitoring observations carried out between October 24 and November 2, 2007 (which had a total exposure time of $\sim 11.1~$ksec). At this time, the source activity was lower again; it displayed a 2-10 keV luminosity of $\sim 1.4 \times 10^{33}~\mathrm{erg~s}^{-1}$. 

It is possible that XMM J174457-2850.3 did not reach a luminosity exceeding $\sim 10^{34}~\mathrm{erg~s}^{-1}$ during the above described episode. Nevertheless, the \textit{Swift} monitoring observations show that if the source went into an active state around July-August 2007, the outburst was shorter than $\sim 3$~months, since the source was detected at lower luminosities again in late October 2007. For an average 2-10 keV outburst luminosity of $\sim 10^{36}~\mathrm{erg~s}^{-1}$ (the maximum value ever observed for this source), the 2-10 keV fluency of this possible outburst would have been $\lesssim 7.5 \times 10^{-4}~\mathrm{erg~cm}^{-2}$. If the system was active at a 2-10 keV luminosity of $\sim 10^{34}~\mathrm{erg~s}^{-1}$ for three months, the outburst fluency lowers to $\lesssim 7.5 \times 10^{-6}~\mathrm{erg~cm}^{-2}$ (in the 2-10 keV band).

XMM J174457-2850.3 was detected above its quiescent level several times since its discovery 7 years ago. However, it is unclear whether the source always reaches full outburst with $L_{\mathrm{X}} \sim 10^{36}~\mathrm{erg~s}^{-1}$, or undergoes enhanced levels of activity with luminosities of several times $\sim 10^{33-34}~\mathrm{erg~s}^{-1}$. This makes it difficult to estimate its recurrence time and outburst duration. Since 2001, XMM J174457-2850.3 was detected at outburst luminosities of $\sim 10^{36}~\mathrm{erg~s}^{-1}$ for 5 times. If we roughly assume a typical outburst duration of $\lesssim 3$~months, a lower limit for the duty cycle of the system is $\gtrsim 5 \%$. For an upper limit on the activity of XMM J174457-2850.3, we may crudely estimate that it goes into outburst twice a year, in which case the duty cycle is almost $50\%$. Adopting a typical outburst luminosity of $\sim 10^{36}~\mathrm{erg~s}^{-1}$, then results in an estimated long-term mass-accretion rate of $1 \times 10^{-11}~\mathrm{M}_{\odot}~\mathrm{yr}^{-1} \lesssim \langle \dot{M}_{\mathrm{long}} \rangle \lesssim 1 \times 10^{-10}~\mathrm{M}_{\odot}~\mathrm{yr}^{-1}$ for a neutron star compact primary, or $2 \times 10^{-12}~\mathrm{M}_{\odot}~\mathrm{yr}^{-1} \lesssim \langle \dot{M}_{\mathrm{long}} \rangle \lesssim 2 \times 10^{-11}~\mathrm{M}_{\odot}~\mathrm{yr}^{-1}$  in case it is a black hole.

\section{Discussion}\label{sec:discuss}

\begin{table*}
\begin{threeparttable}[t]
\begin{center}
\caption[]{{Overview of the outburst duration, fluency and estimated long-term averaged accretion rates of the seven transients detected during the \textit{Swift}/GC monitoring campaign of 2006 and 2007.}}
\begin{tabular}{l l l l l l}
\hline
\hline
Source & Year &  $t_{\mathrm{ob}}$\tnote{a}& \~{F }\tnote{b} & $\langle \dot{M}\rangle _{\mathrm{NS}}$\tnote{c} & $\langle \dot{M} \rangle _{\mathrm{BH}}$\tnote{d} \\
\hline
AX J1745.6-2901\tnote{e} & 2006 & $>16$ &$\gtrsim 1.8 \times 10^{-4}$ &  &  \\
 & 2007 & $>78$ &$\gtrsim 6.5 \times 10^{-3}$ &  $\sim (5-15) \times 10^{-11}$ &   \\
CXOGC J174553.5-290124 & 2006 & $>12$ &$\gtrsim 1.6 \times 10^{-5}$ & $\sim (5-13) \times 10^{-13}$ & $\sim (7-18) \times 10^{-14}$ \\
CXOGC J174540.0-290005  & 2006 & 2 & $1.3\times 10^{-5}$ & $ \sim (3-15) \times 10^{-13}$ & $ \sim (4-21) \times 10^{-14}$ \\
Swift J174553.7-290347 & 2006 & 2 & $8.0\times 10^{-6}$ &$\lesssim 1 \times 10^{-12}$ & $\lesssim 2 \times 10^{-13}$ \\
Swift J174622.1-290634 & 2006 &  5 & $5.0\times 10^{-6}$ & $\lesssim 4 \times 10^{-13}$ &  $\lesssim 6 \times 10^{-14}$\\
GRS 1741.9-2853\tnote{e} & 2006 & 1 & $2.6 \times 10^{-6}$ &  &  \\
 & 2007 & $>13$ & $\gtrsim 1.4 \times 10^{-3}$ & $\sim (2-8) \times 10^{-11}$ &  \\
XMM J174457-2850.3 & 2007 & $<12$ & $\lesssim 7.5 \times 10^{-4}$ &  $\sim (1-10) \times 10^{-11}$ &  $\sim (2-20) \times 10^{-12}$\\
\hline
\end{tabular}
\label{tab:mdot}
\begin{tablenotes}
\item[a]Outburst duration in weeks.
\item[b]Outburst fluency in units of $\mathrm{erg~cm}^{-2}$ in the 2-10 keV energy band, which was calculated by multiplying the mean unabsorbed outburst flux by the outburst duration.
\item[c]Estimated long-term averaged accretion rate ($\mathrm{M}_{\odot}~\mathrm{yr}^{-1}$) for a neutron star with $M=1.4~\mathrm{M_{\odot}}$ and $R=10~\mathrm{km}$
\item[d]Estimated long-term averaged accretion rate ($\mathrm{M}_{\odot}~\mathrm{yr}^{-1}$) for a black hole with $M=10~\mathrm{M_{\odot}}$ and $R=30~\mathrm{km}$.
\item[e]AX J1745.6-2901 and GRS 1741.9-2853 both display type-I X-ray bursts and are thus confirmed neutron star systems.
\end{tablenotes}
\end{center}
\label{tab:obdurations}
\end{threeparttable}
\end{table*}

We have presented the spectral analysis of seven X-ray transients that were found to be active during a monitoring campaign of the field around $\mathrm{Sgr~A}^{\ast}$ using \textit{Swift}/XRT, carried out in 2006-2007. Two new transients were discovered (Swift J174622.1-290634 and Swift J174553.7-290347) and renewed activity from five known sources was observed (AX J1745.6-2901, CXOGC J174553.5-290124, CXOGC J174540.0-290005, GRS 1741.9-2853 and XMM J174457-2850.3). 
Adopting source distances of 8 kpc, we can infer peak luminosities in the range of $\sim 1 \times 10^{34} - 6 \times 10^{36}~\mathrm{erg~s}^{-1}$ in the 2-10 keV energy band. The two transients AX J1745.6-2901 and GRS 1741.9-2853 are hybrid systems, that display very-faint outbursts with 2-10 keV peak luminosities of $L_{\mathrm{X}}<10^{36}~\mathrm{erg~s}^{-1}$, as well as outbursts with luminosities in the range of $10^{36-37}~\mathrm{erg~s}^{-1}$, which are classified as faint. The other five systems display 2-10 keV peak luminosities of $10^{34-36}~\mathrm{erg~s}^{-1}$, i.e., in the very-faint regime. We have observed a large variation in spectral properties, outburst luminosities and outburst durations (see Tables~\ref{tab:spectra}~and~\ref{tab:mdot}). In that respect, the subluminous transients are not different from the well-known bright systems.

\subsection{The nature of the detected transients}
AX J1745.6-2901 and GRS 1741.9-2853 are both known X-ray bursters, which makes it very likely that these are neutron stars in LMXBs, since type-I X-ray bursts have never been detected from high-mass X-ray binaries. For AX J1745.6-2901 a LMXB nature is confirmed by its orbital period of 8.4 hours. The nature of the remaining five transients is unknown. However, XMM J174457-2850.3, CXOGC J174553.5-290124 and CXOGC J174540.0-290005, all have been in outburst more than once in the past decade. This likely rules out a white dwarf accretor, since recurrent novae display outburst cycles of decades rather than years. This suggests a neutron star or black hole nature for XMM J174457-2850.3, CXOGC J174553.5-290124 and CXOGC J174540.0-290005. 
For CXOGC J174540.0-290005, observations reported by \citet{wang06_atel935} could not detect a near-IR counterpart, while the observations would have detected a main sequence star down to spectral type B5. This suggests that if CXOGC J174540.0-290005 is an X-ray binary, it is likely a LMXB.

The two new transients Swift J174553.7-290347 and Swift J174622.1-290634 were observed at 2-10 keV peak luminosities of $2.0 \times 10^{35}~\mathrm{erg~s}^{-1}$ and $7.0 \times 10^{34}~\mathrm{erg~s}^{-1}$. Although such luminosities are quite uncommon for white dwarf systems, \citet[][]{mukai08} showed a few examples of classical novae that reach peak values of several times  $10^{34-35}~\mathrm{erg~s}^{-1}$. Thus, in absence of other outbursts from Swift J174553.7-290347 and Swift J174622.1-290634, we cannot exclude the possibility that these two systems harbor accreting white dwarfs.

\subsection{subluminous X-ray transients in quiescence}\label{quiescence}
The quiescent luminosity of X-ray transients sometimes holds clues to the nature of the system. The \textit{ASCA} burster AX J1745.6-290 is very likely associated with CXOGC J174535.6-290133, which was detected several times with \textit{Chandra} at a level of a few times $10^{32}~\mathrm{erg~s}^{-1}$ (see Sect.~\ref{subsec:bron1}). This is consistent with the neutron star nature of AX J1745.6-290, since black hole systems with an orbital period of $\sim$~8~hours are significantly fainter \citep[e.g.,][]{narayan97,menou99,lasota07}. 
GRS 1741.9-2853 is also a confirmed neutron star system and displays a similar quiescent level of $\sim 10^{32}~\mathrm{erg~s}^{-1}$ \citep[2-8 keV,][]{muno03_grs}.
 
The possible quiescent counterpart of the new subluminous X-ray transient Swift J174553.7-290347, the \textit{Chandra}-detected X-ray source CXOGC J174553.8-290346, displays a 2-10 keV X-ray luminosity of $\sim 7 \times 10^{31}-4 \times 10^{32}~\mathrm{erg~s}^{-1}$, depending on the assumed spectral model (see Sect.~\ref{subsec:bron4}). 
The quiescent luminosity of XMM J174457-2850.3 is also in this regime; $\sim 10^{32}~\mathrm{erg~s}^{-1}$ \citep[2-10 keV;][]{sakano05}. If Swift J174553.7-290347 and XMM J174457-2850.3 are X-ray binaries, their quiescent luminosities are relatively high and might point towards a neutron star nature \citep[e.g.,][]{lasota07}, athough the orbital period of both these systems is unknown. 
We note that the absorption towards our transients is very high ($> 6 \times 10^{22}~\mathrm{cm}^{-2}$). Therefore, any thermal emission from the neutron star surface cannot be observed and we can only detect contributions from a powerlaw component, which is frequently observed for neutron stars at similarly low quiescent luminosities \citep[e.g.,][]{jonker07_eos}.

Two other transients, CXOGC J174553.5-290124 and CXOGC J174540.0-290005, were not detected in quiescence, but have upper limits on their luminosities of $< 9 \times 10^{30}~\mathrm{erg~s}^{-1}$ and $< 4 \times 10^{31}~\mathrm{erg~s}^{-1}$ respectively \citep[2-8 keV;][]{muno05_apj622}. Such low quiescent luminosities are more common for black hole X-ray binaries than for neutron star systems \citetext{e.g., \citealp{garcia01,lasota07}, but see \citealp{jonker06,jonker07}}.

\subsection{The outbursts of subluminous X-ray transients}\label{subsec:outbursts}
The disk instability model \citep[e.g.,][]{king98, dubus99, lasota01} provides a framework to describe the outburst cycles of transient LMXBs. However, it is unclear why some X-ray transients, such as the ones discussed here, undergo outbursts with very low peak luminosities. AX J1745.6-2901 has a known orbital period of 8.4 hours, which allows for a maximum luminosity of $\sim 2 \times 10^{38}~\mathrm{erg~s}^{-1}$ \citep[for a hydrogen-dominated disk and a neutron star mass of $1.4~\mathrm{M_{\odot}}$;][]{lasota07}. Yet, its observed peak luminosity is over an order of magnitude lower (see Table~\ref{tab:spectra}). Since AX J1745.6-2901 displays eclipses, we must look at the system at high inclination. For several eclipsing X-ray binaries observations suggest that these are intrinsically bright but appear faint because the bright center of the system is blocked by the outer edge of the disk and the corona \citep[e.g.,][]{parmar00, kallman03,muno05_apj633}. This may also be the case for AX J1745.6-2901, for which \citet{maeda1996} derived an inclination angle of $i \sim 70^{\circ}$. To include inclination effects, the observed X-ray luminosity should be corrected by a factor $\xi_{p}$, which relates to the inclination, $i$, as $\xi_{p}^{-1}= 2 |\cos i|$ \citep{fujimoto88,lapidus85}. In 2007, AX J1745.6-2901 displayed a 2-10 keV peak luminosity of $6.1 \times 10^{36}~\mathrm{erg~s}^{-1}$, which corrects to $9.2 \times 10^{36}~\mathrm{erg~s}^{-1}$ for the suggested inclination of $i \sim 70^{\circ}$ ($\xi_{p} \sim 1.5$). 

It is thus conceivable that AX J1745.6-2901 is a bright X-ray transient that is obscured due to line of sight effects, although it would still seem to be at the lower end of the luminosity range for bright systems (peak luminosities of $\sim 10^{37-39}~\mathrm{erg~s}^{-1}$ in the 2-10 keV energy band). For comparison, the quasi-persistent neutron star system MXB 1659-29 has an orbital period of 7.1 hours \citep{mxb1659_eclipses}, which is close that of AX J1745.6-2901. However, MXB 1659-29 displays an average 2-10 keV outburst luminosity of $7 \times 10^{36} ~(d/10 ~\mathrm{kpc})~\mathrm{erg~s}^{-1}$ \citep[][]{oosterbroek01_mxb,sidoli01_mxb}, which is about a factor of 4 higher than the average 2-10 keV outburst luminosity observed for AX J1745.6-2901 in 2007; $1.6 \times 10^{36} ~\mathrm{erg~s}^{-1}$. Possibly, the inclination of AX J1745.6-2901 is somewhat higher than the $i \sim 70^{\circ}$ suggested by \citet{maeda1996}. AX J1745.6-2901 might have a subluminous appearance due to line of sight effects, but it is important to note that statistical arguments show that such effects cannot account for the entire population of subluminous X-ray transients, and that most systems must have low intrinsic luminosities \citep[see the discussion of][]{wijn06_monit}.

Although taking into account inclination effects potentially pushes AX J1745.6-2901 into the regime of bright X-ray transients, this does not provide an explanation for the peculiar outburst behavior of the source. As discussed in Sect.~\ref{subsec:bron1}, the system was likely in quiescence for several years before it was seen active in 2006 for more than 4 months. At that time, the source reached a peak luminosity of $9.2 \times 10^{35}~\mathrm{erg~s}^{-1}$, which would classify the system as very-faint. However, after several months of quiescence (see Fig.~\ref{fig:lc}), the source reappeared displaying a peak luminosity of $6.1 \times 10^{36}~\mathrm{erg~s}^{-1}$ (i.e., in the faint regime) and remained active for over 1.5 years (see Sect.~\ref{subsec:bron1}). Thus, the outburst observed in 2006 was subluminous by about a factor of 6 compared to the 2007 outburst, yet the system maintained this low luminosity for months. It is unclear if this behavior can be explained in terms of a disk instability model. In 1993 and 1994, different outburst luminosities of $2 \times 10^{35}~\mathrm{erg~s}^{-1}$ and $9 \times 10^{35}~\mathrm{erg~s}^{-1}$ were reported for AX J1745.6-2901 \citep[3-10 keV,][]{maeda1996}. This is on the same time scale as the discussed \textit{Swift} detections, suggesting that the behavior observed in 2006 and 2007 could be typical for the source. 

GRS 1741.9-2853 also displayed two separate outbursts with very different characteristics in terms of peak luminosity and outburst duration during the \textit{Swift}/XRT monitoring observations. A short, $\sim1-$week outburst was detected in 2006, which had a 2-10 keV peak luminosity of $9.2 \times 10^{34}~\mathrm{erg~s}^{-1}$. A few months later, the source exhibited a much longer ($\gtrsim 13$~weeks) outburst, that reached a peak luminosity of $2.0 \times 10^{36}~\mathrm{erg~s}^{-1}$ (2-10 keV). Possibly, the short 2006 outburst of GRS 1741.9-2853 was an X-ray precursor for the 2007 outburst. Such behavior is observed for several bright X-ray transients \citep[both neutron star and black hole systems, see][and references therein]{chen97}. Both Swift J174553.7-290347 and CXOGC J174540.0-290005 displayed short, $\sim$ 2-week outbursts that had an average luminosity of a few times $10^{34}~\mathrm{erg~s}^{-1}$. This kind of activity resembles the small accretion outburst of GRS 1741.9-2853 in 2006 (see Fig.~\ref{fig:lc}), but for these two systems no longer outbursts have been observed. XMM J174457-2850.3 seems to undergo X-ray activity at different luminosity levels as well (see Sect.~\ref{subsec:bron7}). It is unclear what causes the varying accretion luminosities. However, this phenomenon is also observed for bright X-ray transients and is thus not restricted to the subluminous systems discussed here.

Current disk instability models do not provide an obvious explanation for accretion outbursts that last several years, rather than the usual weeks to months, such as observed for AX J1745.6-2901. A few bright systems are known to undergo quasi-persistent outbursts \citep[see e.g.,][]{wijnands04_quasip}. There are also two X-ray transients that exhibit prolonged outbursts at low luminosities. XMMU J174716.1-281048 has likely been continuously active since its initial discovery in 2003, displaying a typical  2-10 keV luminosity of a few times $10^{34}~\mathrm{erg~s}^{-1}$  \citep[e.g.,][]{delsanto07,degenaar07_xmmsource}. Furthermore, AX J1754.2-2754 recently made a transition to quiescence \citep{bassa08}, after exhibiting an accretion outburst with a 2-10 keV luminosity of several times $10^{34-35}~\mathrm{erg~s}^{-1}$, which likely lasted for 7-8 years \citep[][]{sakano02,delsanto07_ascabron,chelovekov07_ascabron}. This source was again found active in July 2008 \citep{jonker08}. The detection of type-I X-ray bursts identifies both these systems as neutron star LMXBs, just like AX J1745.6-2901.

\subsection{X-ray bursts from subluminous X-ray transients}
The properties of type-I X-ray bursts are set by the conditions in the flash layer such as the temperature, thickness, hydrogen abundance and the fraction of carbon-nitrogen-oxygen (CNO) elements in the layer \citep[e.g.,][]{fujimoto81,bildsten98,peng2007}. These conditions can vary drastically as the mass-accretion rate onto the neutron star ($\dot{M}$) varies, which results in flashes with different characteristics for different $\dot{M}$ regimes \citep[e.g., ][]{fujimoto81,peng2007,cooper07}.

The \textit{Swift}/XRT monitoring observations of 2006 caught two type-I X-ray bursts from AX J1745.6-2901 (see Sect.~\ref{subsec:bron1}). The average 2-10 keV luminosity of the 2006 outburst was $3.9 \times 10^{35}~\mathrm{erg~s}^{-1}$, from which we can estimate an instantaneous mass accretion rate onto the neutron star of $\sim1 \times 10^{-10}~\mathrm{M_{\odot}~yr}^{-1}$. If we include a correction factor to account for inclination effects, as discussed in Sect.~\ref{subsec:outbursts}, this value increases to $\sim1.5 \times 10^{-10}~\mathrm{M_{\odot}~yr}^{-1}$. The bursts had a duration of $50-60~\mathrm{seconds}$ (see Fig.~4), which suggests triggering in a mixed hydrogen/helium environment. This is in line with the classical predictions for the estimated mass-accretion rate \citep[e.g.,][]{fujimoto81}. 

We discussed in Sect.~\ref{subsec:bron1}, that the type-I X-ray bursts observed from AX J1745.6-2901 have 0.01-100 keV peak luminosities of $\sim 10^{38}~\mathrm{erg~s}^{-1}$, close to the Eddington limit of a neutron star. These peak values should also be corrected for the inclination effects discussed in Sect.~\ref{subsec:outbursts}. However, the X-ray burst (originating form the neutron star surface) and the outburst emission (emerging from the accretion disk) are attributed to geometrically different regions and may therefore have different degrees of isotropy \citep{fujimoto88,lapidus85}. Although, the X-ray burst emission will be partly intercepted and re-radiated by the accretion disk, it was shown that the degree of anisotropy is less than for the emission coming from the accretion disk \citep{fujimoto88,lapidus85}. Inclination effects are expected to reduce the X-ray burst emission by a factor $\xi_{b}^{-1}= 0.5 + |\cos i |$ \citep{fujimoto88,lapidus85}. For the suggested inclination of $i=70^{\circ}$ \citep{maeda1996}, we thus obtain a correction factor of $\xi_{b}=1.2$. This implies peak luminosities for the type-I X-ray bursts observed from AX J1745.6-2901 on June 3 and June 14, 2006, of  $(1.2-1.6) \times 10^{38}~\mathrm{erg~s}^{-1}$ and $(0.62-1.3) \times 10^{38}~\mathrm{erg~s}^{-1}$ (0.01-100 keV) respectively.  This is below, but close, to the Eddington luminosity for a neutron star \citep[$2.0 \times 10^{38}\mathrm{~erg~s}^{-1}$ for a hydrogen-rich and $3.8 \times 10^{38}\mathrm{~erg~s}^{-1}$ for a hydrogen-poor photosphere; e.g.,][]{kuulkers03_xrb}.

\subsection{Long-term average accretion rates}\label{mdot_estimate}
Presuming that the detected transients are accreting systems, we attempted to estimate their long-term time-averaged accretion rates using the method and described in Sect.~\ref{subsec:accrates}. We explored the scenarios of both neutron star and black hole accretors (except for AX J1745.6-2901 and GRS 1741.9-2853, since these are confirmed neutron star systems), which resulted in the estimated long-term mass-accretion rates\footnote{Note the caveat mentioned for the black hole cases in Sect.~\ref{subsec:accrates}.} listed in Table~\ref{tab:obdurations}. 
The two confirmed neutron star systems, AX J1745.6-290 and GRS 1741.9-2853 have the highest estimated accretion rates of the seven discussed transients ($\sim 10^{-11}-10^{-10}~\mathrm{M_{\odot}~yr}^{-1}$). This arises from the fact that GRS 1741.9-2853 is active quite regularly and AX J1745.6-290 can be in outburst for a very long time (over 1.5~years). The regime estimated for these two sources can be well explained within current LMXB evolution models. The same is likely true for XMM J174457-2850.3, which was active several times since its discovery in 2001 and has an estimated long-term mass-accretion rate of $\gtrsim 10^{-11}~\mathrm{M_{\odot}~yr}^{-1}$.

The estimates for the remaining four systems, CXOGC J174553.5-290124, CXOGC J174540.0-290005, Swift J174553.7-290347 and Swift J174622.1-290634 are much lower; $\lesssim 10^{-12}~\mathrm{M_{\odot}~yr}^{-1}$ for accreting neutron stars and even an order of magnitude lower for black hole X-ray binaries, $\lesssim 10^{-13}~\mathrm{M_{\odot}~yr}^{-1}$ (see Table~\ref{tab:obdurations}).
Comparing our results with a theoretical toy-model of \citet{king_wijn06}, who explored the mechanism of Roche-lobe overflow at low accretion rates, suggests that if these transients are LMXBs, their low time-averaged mass-accretion rates might pose difficulties explaining their existence, without invoking exotic scenarios such as accretion from a planetary donor or an intermediate mass black hole as the accreting primary \citep{king_wijn06}. These are thus interesting systems to track and monitor in the future. 

Apart from evolutionary scenarios and line-of-sight effects, there other possible explanations for the subluminous X-ray appearance of these transients. For example, 
in particular for the systems containing a black hole, the liberated accretion power may not be primarily dissipated as X-rays but rather via radiatively inefficient flows \citep[e.g.,][]{blandford99,fender03,narayan}. 
Furthermore, in neutron star systems the "propeller mechanism" can possibly operate, so that only a small fraction of the mass transferred from the donor can be accreted onto the neutron star~\citep[e.g.,][]{illarionov1975,alpar2001,romanova2005}. 

The discussed examples of AX J1745.6-2901, GRS 1741.9-2853 and XMM J174457-2850.3 illustrate that X-ray transients can display different behavior in terms of peak luminosity, outburst duration and recurrence time from year to year. It is currently not understood whether these variations should be interpreted as, e.g., being due to changes in the mass-transfer rate from the donor star or as the result of instabilities in the accretion disk. Such issues need to be resolved before we can fully comprehend the nature of subluminous X-ray transients.

\section*{Acknowledgments}
The auhors thank Anna Watts for commenting on an early version of this manuscript and an anonymous referee for giving valuable suggestions. We acknowledge the use of public data from the \textit{Swift} data archive. This work was supported by the Netherlands Organization for Scientific Research (NWO).

\bibliographystyle{aa}
\bibliography{0654}
\end{document}